\documentclass[11pt]{article}
\usepackage{amsmath,amssymb, amsfonts, bm}
\usepackage{rotating}
\usepackage{caption}

\def\eo{\overset{_{\phantom{.}\circ}}{e}{}}

\def\go{\overset{_{\phantom{.}\circ}}{g}{}}

\def\Do{\overset{_{\phantom{.}\circ}}{D}{}}

\def\etao{\overset{_{\phantom{.}\circ}}{\eta}{}}
\def\zetao{\overset{_{\phantom{.}\circ}}{\zeta}{}}

\newcommand{\fr}{\mathfrak{f}_{\scriptscriptstyle{F \hspace{-0.8mm} R}}}

\newcommand{\cV}{{\cal V}}
\newcommand{\cM}{{\cal M}}
\newcommand{\cN}{{\cal N}}

\newcommand{\cB}{{\cal B}}
\newcommand{\cD}{{\cal D}}
\newcommand{\cG}{{\cal G}}
\newcommand{\cH}{{\cal H}}
\newcommand{\cQ}{{\cal Q}}
\newcommand{\cO}{{\cal O}}
\newcommand{\cP}{{\cal P}}

\newcommand{\cL}{{\cal L}}

\newcommand{\rX}{{\rm X}}
\newcommand{\rY}{{\rm Y}}

\begin{document}

\begin{titlepage}

\hfill  DAMTP-2015-41 

\vspace{2.5cm}
\begin{center}

{{\LARGE  \bf  Consistent 4-form fluxes for \\[3mm] maximal supergravity}} \\

\vskip 1.5cm {Hadi Godazgar$^{\star}$, Mahdi Godazgar$^{\star}$, Olaf Kr\"uger$^{\dagger}$
and Hermann Nicolai$^{\dagger}$ }
\\
{\vskip 0.8cm
$^{\star}$DAMTP, Centre for Mathematical Sciences,\\
University of Cambridge,\\
Wilberforce Road, Cambridge, \\ CB3 0WA, UK\\
\vskip 0.2cm
$^{\dagger}$Max-Planck-Institut f\"{u}r Gravitationsphysik, \\
Albert-Einstein-Institut,\\
Am M\"{u}hlenberg 1, D-14476 Potsdam, Germany}
{\vskip 0.8cm
H.M.Godazgar@damtp.cam.ac.uk, M.M.Godazgar@damtp.cam.ac.uk, \\
\vskip 0.08cm

Olaf.Krueger@aei.mpg.de, Hermann.Nicolai@aei.mpg.de}
\end{center}

\vskip 0.35cm

\begin{center}
October 25, 2015
\end{center}

\noindent

\vskip 1.2cm

\begin{abstract}

{\noindent
We derive new ans\"atze for the 4-form field strength of $D=11$ supergravity corresponding to uplifts of four-dimensional maximal gauged supergravity. In particular, the ans\"atze directly yield the components of the 4-form field strength in terms of the scalars and vectors of the four-dimensional maximal gauged supergravity---in this way they provide an \emph{explicit} uplift of \emph{all} four-dimensional consistent truncations of $D=11$ supergravity. The new ans\"atze provide a substantially simpler method for uplifting $d=4$ flows compared to the previously available method using the 3-form and 6-form potential ans\"atze. The ansatz for the Freund-Rubin term allows us to conjecture a `master formula' for the latter in terms of the scalar potential of $d=4$ gauged supergravity and its first derivative. We also resolve a long-standing puzzle concerning the antisymmetry of the flux obtained from uplift ans\"atze.}

\end{abstract}

\end{titlepage}

\section{Introduction}
\label{sec:introduction}

Establishing a formal, consistent relation between a higher-dimensional theory and a lower dimensional one is, in general, a
challenging problem due to the highly non-linear nature of reductions.
Given some (super-)gravity model in $D$ dimensions, consider a ground
state solution
\begin{equation}\label{MD}
  \cM_{D} \, = \, \cM_4 \times \cM_{D-4} 
\end{equation}
corresponding to a compactification from $D$ to four dimensions. The
fields of the theory are then expanded linearly around this ground
state according to
\begin{equation}\label{Phi}
  \Phi(x,y) \, = \Phi_{0} (x,y) + \, \sum_n \Phi^{(n)}(x) Y^{(n)}(y),
\end{equation}
where we collectively denote the value of the fields (metric and form fields) at the 
ground state by $ \Phi_{0} (x,y)$. 
Here, $x^\mu$ and $y^m$, respectively, are four-dimensional `external'
and ($D-4$)-dimensional `internal' coordinates on $\cM_4$ and
$\cM_{D-4}$. The $Y^{(n)}(y)$ are the eigenmodes of certain
differential operators on the internal space giving rise to an
infinite tower of Kaluza-Klein modes. Restricting to the zero-mass
eigenmodes gives the low energy physics. The linearised expansion
(\ref{Phi}) is sufficient to determine the mass spectrum of the
theory. However, it cannot provide complete information about the
interactions of the low energy theory, and must be modified by
non-linear terms away from an infinitesimal neighborhood of the ground
state. This modification must ensure that any solution of the low
energy theory corresponds to a solution of the higher-dimensional
theory. This is the problem of {\em Kaluza-Klein consistency}\,: given
any solution of the full non-linear field equations in four dimensions
one must seek a corresponding expression for $\Phi(x,y)$ that solves
the full higher-dimensional field equations also away from $\Phi_0(x,y)$, thereby 
arriving at a consistent embedding of this solution into the higher-dimensional
theory.

In fact, there are very few examples 
where such a program has been successfully completed. Beyond the task
of establishing the consistency of the truncation, it is a major
challenge to present {\em explicit} non-linear
ans\"atze\footnote{Here, we use the word ``ansatz'' in the sense of an
  approach or prescription rather than a guess.} for uplifting
solutions of the lower-dimensional theory to solutions of the higher-dimensional one. Among the known examples the most intricate and
technically demanding concerns the maximally supersymmetric $D=11$
supergravity and reductions thereof to maximal gauged supergravity
theories in four dimensions, corresponding to the ground state
\begin{equation}\label{M7}
  \cM_{11} \, = \,  {\rm AdS}_4 \times \cM_{7}.
\end{equation}
For this theory the complete non-linear ans\"atze have recently been identified 
in Refs.~\cite{Godazgar:2013dma,Godazgar:2013pfa}, building on the results of Refs.~\cite{deWit:1986iy,deWit:1984nz,dWN13} and using the formalism developed in Ref.~\cite{deWit:1986mz}. 
The basic tool that facilitates this result
is the reformulation of the $D=11$ supergravity theory
\cite{Cremmer:1978km} such that essential features of maximal gauged
supergravity theories, classified by the covariant embedding formalism \cite{Nicolai:2000sc,
  Nicolai:2001sv, deWit:2002vt, deWit:2003hr, deWit:2007mt}, in four dimensions become manifest. At its
heart lies the E$_{7(7)}$/SU(8) duality symmetry \cite{cremmerjulia,
  so(8)}, which is obtained in the toroidal reduction from $D=11$
supergravity to four-dimensional ungauged maximal supergravity. 
An important aspect of the formalism developed in
Ref.~\cite{Godazgar:2013dma} is the role of the 6-form potential,
which is dual to the 3-form potential of $D=11$ supergravity.  Ref.~\cite{Godazgar:2013pfa} (see also
Refs.~\cite{deWit:1984nz, dWN13, Godazgar:2013nma}) derives full, explicit uplift ans\"atze for
SO(8) gauged maximal supergravity \cite{deWit:1982ig}\footnote{It is
  known that the recently discovered family of SO(8) gauged
  supergravity theories \cite{Dall'Agata:2012bb, Dall'Agata:2014ita}
  cannot be obtained from a consistent reduction of $D=11$
  supergravity \cite{Godazgar:2013pfa, Godazgar:2013oba, Lee:2015xga}
  (see also Ref.~\cite{dWN13}). Therefore, they fall outside the scope
  of this paper.}, which is a consistent
truncation~\cite{deWit:1986iy, NP} of $D=11$ supergravity on a
seven-sphere \cite{Freund:1980xh, Duff:1983gq}.

The non-linear ans\"atze for the internal metric and internal
components of the form fields were obtained by an analysis of the
supersymmetry variations of $D=11$ supergravity. In
particular, the supersymmetry transformation of those components of
the fields that we identify with the vectors in a reduction take the
same form as the supersymmetry transformation of the vectors in four
dimensions, \emph{viz}.\ both are given by components of a 56-bein
multiplied by a particular combination of fermions. Hence given a
linear ansatz for the vectors, one can relate the 56-bein in eleven dimensions to
the four-dimensional one. Since these 56-beine are parametrised by the
$d=4$ scalars and the internal components of the $D=11$ fields
respectively, one finally obtains a non-linear ansatz that relates the
internal components of the $D=11$ fields to the $d=4$ scalars.

By contrast, the approach in this paper is based on an analysis of the generalised
vielbein postulates (GVPs). These are analogues of the familiar
vielbein postulate in differential geometry for the 56-bein. As in the
simpler case of the vielbein postulate, the GVPs express the
derivative of the 56-bein in terms of objects that transform as
connections with respect to SU(8) transformations or E$_{7(7)}$
generalised diffeomorphisms \cite{Coimbra:2011ky}. The GVPs, used in
this paper, are found \cite{Godazgar:2013dma} by expressing the
56-bein in a GL(7) decomposition (in terms of the components of the
$D=11$ fields) and by packaging its derivative in terms of generalised
connections.
This alternative method for finding non-linear ans\"atze (see
Ref.~\cite{deWit:1986iy}), centres on the fact that the generalised
connections are parametrised by, in particular, components of the
4-form field strength. Therefore, by projecting onto various
components of the GVPs using the 56-bein we are able to extract
non-linear ans\"atze for components of the 4-form field strength.

One main result of this paper is the embedding formula for the Freund-Rubin parameter
$\fr(x,y)$ in terms of four-dimensional fields. The latter is generally and independently of
the equations of motion defined by \cite{Freund:1980xh}
\begin{equation} \label{4dF} 
  F_{\mu \nu \rho \sigma} (x,y) = i\,
  \fr(x,y)\, \etao_{\mu \nu \rho \sigma},
\end{equation}
where $\etao_{\mu \nu \rho \sigma}$ is the volume form in four dimensions.
The choice of terminology reflects the fact that $\fr$ is a constant for Freund-Rubin 
compactifications characterised by (\ref{M7}). On the basis of its observed structure
for several examples (worked out in section \ref{sec:testing-non-linear} and 
appendices \ref{app:freund-rubin-so3t} and \ref{app:freund-rubin-su4}) we 
conjecture the following {\em master formula}
\begin{equation}
  \label{eq:conjecture-intro}
  \fr (x,y) = -\frac{m_7}{\sqrt{2}g^2}
  \left(
    V(x)  - \frac{g^2}{24}
    \left(
      Q^{ijkl}(x) \hat \Sigma_{ijkl}(x,y) + \mathrm{h.c.}
    \right)
  \right),
\end{equation}
where $m_7$ is the inverse radius of the round $S^7$. Here, $V$ is the full 
scalar potential of gauged maximal $N=8$ supergravity with gauge 
coupling constant $g$. $Q^{ijkl}(x)$ is the first derivative of 
the potential in an SU(8) covariant `frame' on the E$_{7(7)}$/SU(8)
coset manifold (see Ref.~\cite{deWit:1983gs} and section
\ref{sec:state-conjecture} for details), and $ \hat \Sigma_{ijkl}$ is
the $x$- and $y$-dependent complex selfdual tensor defined in
Eqn.~(\ref{Sigmahat}) in section \ref{sec:state-conjecture}.
Stationary points of the potential are therefore characterised by the
requirement that $Q^{ijkl}$ be complex {\em anti}-selfdual; at such
points the $y$-dependence drops out. We perform several very
non-trivial checks of the formula (\ref{eq:conjecture-intro}), but
leave a general proof 
for later work.

The master formula (\ref{eq:conjecture-intro}) provides a concrete example
of how a higher-dimensional field $\Phi(x,y)$ is consistently deformed away
from the ground state solution $\Phi_0(x,y)$. At the same time it illustrates very
explicitly that the consistency of the truncation can
only be achieved {\em on-shell}, that is, when the equations of motion
are obeyed. Away from the solution of the equations of motion, the
Freund-Rubin term exhibits an irremovable and manifest 
$y$-dependence.~\footnote{Nevertheless, in the general $S^7$
  truncation, a residual $y$-dependence of the Freund-Rubin term for
  {\em non-stationary} solutions can be  consistent if other
  components of the 4-form field strength also contribute.  Consistency is then
  achieved because {\em on-shell} the $y$-dependence of the latter cancels the 
  residual $y$-dependence of $\fr$ in such a way that  all these terms 
  combine to sum up to a $y$-independent right-hand side for the $d=4$ Ricci tensor.}
The same holds true for other components of the $D=11$ fields, as well
as for more complicated solutions of the full $S^7$ truncation with
$x$-dependence. As we already pointed out in our previous work, this
is in marked contrast to the AdS$_7 \times S^4$ compactification of
$D=11$ supergravity \cite{Nastase:1999cb, Nastase:1999kf} where there
exist consistent non-linear ans\"atze that also hold off-shell. The reason is
that in the latter case the scalar field content is directly obtained
without the need to dualise form fields.

Finally, our non-linear ansatz for the internal components $F_{mnpq}$ of the
4-form field strength settles an issue that had been
left unresolved in Ref.~\cite{deWit:1986iy}, which also tried to exploit the idea of 
projecting out the 4-form field strengths from the generalised non-metricity.
The construction could not be completed there because
only part of the generalised vielbein was known; furthermore, as shown much later
in Ref.~\cite{NP}, the ansatz as given in Ref.~\cite{deWit:1986iy}
yields a tensor that is not totally antisymmetric. 
We also use the fermion supersymmetry transformations to find an
ansatz for the $F_{\mu \nu mn}$ component of the 4-form field
strength. With these new ans\"atze, the uplift of flows ($x$-dependent
solutions) to $D=11$ becomes technically relatively straighforward.

\section{Preliminaries}
\label{sec:summary}

A (bosonic) solution of four-dimensional maximal gauged supergravity
is specified by the following bosonic field content:
\begin{align}
  &\text{a vierbein}: \hspace{17mm} e_{\mu}{}^{\alpha}(x), \notag \\[2mm]
  &\text{28 vector fields}: \qquad
  A_{\mu}{}^{IJ}(x), \label{4dfieldcon} \notag \\[2mm]
  &\text{70 scalars}: \hspace{17mm} \hat{\mathcal{V}}(x) =
  \begin{pmatrix}
    u_{ij}{}^{IJ}(x) & v_{ij\;IJ}(x) \\[1mm] v^{ij\;IJ}(x) &
    u^{ij}{}_{IJ}(x) 
  \end{pmatrix},
\end{align}
where the bivector indices $IJ$ denote the \textbf{28} of SL(8,
$\mathbb{R}$). The 28 `electric' vector fields $A_\mu{}^{IJ}$ should
really be thought of as belonging to a \textbf{56} of E$_{7(7)}$,
denoted by $A_{\mu}{}^{\cM}$. In the ungauged theory, the other 28
`magnetic' vectors $A_{\mu\,IJ}$ are obtained by dualising the
original 28 vectors $A_\mu{}^{IJ}$. The scalars $u_{ij}{}^{IJ}$ and
pseudoscalars $v_{ij\;IJ}$ parametrise a coset element
$\hat{\mathcal{V}}(x) \in $ E$_{7(7)}$/SU(8).

On the other hand, a solution of $D=11$ supergravity is given by the
following bosonic field content:
\begin{align} 
  &\text{an elfbein}: \hspace{32.5mm} E_{M}{}^{A}(x,y), \notag \\[2mm]
  &\text{a 3-form potential}: \qquad \hspace{10.5mm}
  A_{MNP}(x,y) \label{11dfieldcon} \\[2mm]
  &\big(\text{or 4-form field strength}: \qquad F_{MNPQ}(x,y) = 24
  \, \partial_{[M}A_{NPQ]}\big), \notag
\end{align}
where $y^{m}$ now are seven-dimensional coordinates.  

An uplift of a four-dimensional solution \eqref{4dfieldcon} to $D=11$
supergravity is a solution of the ${D=11}$ equations of motion,
specified by \eqref{11dfieldcon} that is determined purely by
the four-dimensional field content \eqref{4dfieldcon} and the internal
geometry of $\cM_{7}$ relevant to the reduction; in the case of SO(8)
gauged supergravity this is the seven-sphere $S^7$. Decomposing the
$D=11$ fields in a 4+7 split and interpreting them as
four-dimensional fields based on their index structure gives:
\begin{align}\label{11dfieldcons} 
  &\text{a vierbein}: \hspace{17mm} e_{\mu}{}^{\alpha}(x,y), \notag \\[1mm]
  &\text{28 vector fields}: \qquad B_{\mu}{}^{m}(x,y),\quad A_{\mu
    mn}(x,y),   \\[1mm]
  &\text{70 scalars}: \hspace{17mm} e_{m}{}^{a}(x,y),\quad \
  A_{mnp}(x,y),\quad A_{\mu\nu m}(x,y)\ \quad (\text{or} \;\;
  A_{m_1\dots m_6}(x,y)). \notag
\end{align}
Modulo a Weyl rescaling the eleven-dimensional ``vierbein'' (the
appropriate 4$\times$4 submatrix of the elfbein) is simply identified
with the four-dimensional one. The 28 vector fields can be augmented
by another set of 21 vectors $A_{\mu m_1\dots m_5}(x,y)$ originating
from the 6-form dual field $A_{M_1\cdots M_6}$. The final seven
vectors required to form a full \textbf{56} of E$_{7(7)}$ correspond
to the seven `dual graviphotons' that have no satisfactory
interpretation within $D=11$ supergravity. Nevertheless, for
convenience, we can add seven extra auxiliary vectors (see e.g.
Ref.~\cite{Godazgar:2013dma} and references therein). In this way we
collectively define a set of vectors $B_{\mu}{}^{\mathcal{M}}$, where indices
$\cM, \cN,...$ label the $\bf{56}$ representation of E$_{7(7)}$. These vectors
are related to the analogous set $A_\mu{}^\mathcal{M}$ in four dimensions
by
\begin{eqnarray} \label{eq:vectans}
  B_{\mu}{}^{\mathcal{M}}(x,y) &=&
  \mathcal{R}^{\mathcal{M}}{}_{\cN}(y)\, A_{\mu}{}^{\cN}(x) 
  \, \equiv \,  \mathcal{R}^\cM{}_{IJ}(y) A_\mu{}^{IJ}(x)\, + \, 
  \mathcal{R}^{\cM \, IJ}(y)  A_{\mu\, IJ}(x)\;  .
\end{eqnarray}
Here, $A_\mu{}^{IJ}$ and $A_{\mu\,IJ}$, respectively, are the 28
electric vectors and the 28 magnetic vectors of $N=8$ supergravity. In
the case of the $S^7$ reduction, $\mathcal{R}^{\mathcal{M}}{}_{\cN}$
is constructed from the Killing spinors $\eta^I$ on $S^7$ and the
6-form volume potential on the round $S^7$, $\zetao_{m_1 \dots m_6}$;
the explicit expressions are given in Ref.~\cite{Godazgar:2013pfa}.
Similarly, the eleven-dimensional ``scalars,'' which collectively
define an E$_{7(7)}$/SU(8) coset element
$\mathcal{V}^{\mathcal{M}}{}_{AB}$ \cite{Godazgar:2013dma} are related
to the four-dimensional scalars via
\begin{equation} \label{eq:scalans}
  \mathcal{V}^{\mathcal{M}}{}_{AB}(x,y) =
  \mathcal{R}^{\cM}{}_{\cN}(y)\, \eta_A^{i}(y)\, \eta_B^j(y)\,
  \hat{\mathcal{V}}^{\mathcal{N}}{}_{ij}(x).
\end{equation}
Here, $\eta_A^{i}$ denote the eight Killing spinors defined on the
internal geometry and $\mathcal{R}^{\mathcal{M}}{}_{\cN}$ is the same matrix as in Eqn.~\eqref{eq:vectans}.

In the case of the $S^7$ reduction and the associated SO(8) gauged
supergravity, the above expressions translate to an uplift ansatz for
the internal metric $g_{mn},$ \cite{deWit:1984nz}, the internal 3-form
potential $A_{mnp}$ \cite{dWN13,Godazgar:2013nma, Godazgar:2013pfa}
and the internal 6-form potential $A_{m_1 \dots m_6}$
\cite{Godazgar:2013pfa}. Furthermore, dualisation of the 6-form
potential gives components of the 3-form potential. All these fields
obtained in this way represent a full constructive solution of the
$D=11$ equations of motion. The two-form fields $A_{\mu \nu m}$ can be
obtained by integration from the other ans\"atze. It is in principle
also possible to deduce a non-linear ansatz directly for $A_{\mu \nu
  m}$ by also comparing the four and eleven-dimensional supersymmetry
transformations. Except that in this case the supersymmetry
transformation of $A_{\mu \nu m}$ will correspond in four dimensions
to the supersymmetry transformation of the {\textbf{133}} two-form
fields, $A_{\mu \nu {\bm{\alpha}}},$ in the tensor hierarchy (see
Ref.~\cite{deWit:2007mt}).

It must be emphasised that the uplift ans\"atze have been derived from
the $D=11$ theory, with the supersymmetry transformations playing a
significant role in the derivation. As such they are robust and need
no further substantiation. However, given the non-trivial nature of
the reduction on the one hand and the remarkably simple form of the
ans\"atze on the other, they have been explicitly verified for a
number of stationary points of the four-dimensional scalar potential
including the SO(7)$^\pm$, G$_2$ and SU(4)$^-$ invariant solutions
\cite{deWit:1984nz, NP, Godazgar:2013nma, Godazgar:2013pfa}.
Furthermore, the metric ansatz has been used extensively in the
literature, in particular in applications to holography (see for
example \cite{Cvetic:2000tb, Corrado:2001nv}). The full uplift
ans\"atze have allowed for a study of more complicated upliftings;
including an uplift of the SO(3)$\times$SO(3) invariant solution
\cite{Godazgar:2014eza} and for the first time a full uplift of a flow
to eleven dimensions \cite{Pilch:2015vha}.

In this paper, we explore the possibility of expressing some of the
uplift ans\"atze in even simpler terms, with particular focus on the
Freund-Rubin term (\ref{4dF}) that plays a central role in compactifications of
$D=11$ supergravity. To illustrate the simplicity of our final
formula (\ref{eq:conjecture-intro}) recall the  
duality relation in eleven dimensions between the 4-form field strength 
and its 7-form dual, which implies that 
the Freund-Rubin term can also be expressed (in form language) as
\begin{equation} \label{frUA} 
  \fr = \star_{(7)} \left( \mathrm{d_{(7)}} A_6 - A_3 \wedge F_4
  \right), 
\end{equation}
where all fields above take components along the internal directions.
Hence, a direct derivation of the Freund-Rubin term from the uplift
ans\"atze of Ref.~\cite{Godazgar:2013pfa} would require the associated
expressions for $A_{mnp}$ and $A_{m_1 \dots m_6}$. Although Eqn.~\eqref{frUA}
and the uplift ans\"atze for $A_{mnp}$ and $A_{m_1 \dots m_6}$ are relatively 
simple \footnote{In fact, the ansatz for $A_{m_1 \dots m_6}$ given in
  Ref.~\cite{Godazgar:2013pfa} can be greatly
  simplified: \begin{equation} A_{m_1 \dots m_6} =
    \frac{\sqrt{2}}{16\cdot 5! \, m_7} \go^{pq} \etao_{m_1 \dots m_6 p}
    \Do_{q}(\textup{log}\Delta) - 3 \sqrt{2}\, \zetao_{m_1 \dots
      m_6}. \end{equation}}, in practice the calculations
become rather unwieldy for more non-trivial solutions of the
four-dimensional theory, at least analytically. 
More precisely, the large number of operations required (such as
inverting the metric to find $A_{mnp}$ and $A_{m_1 \dots m_6}$, taking
exterior derivatives and dualising a 7-form) to find what is
ultimately a scalar makes it a rather inconvenient calculation.

Observing that the Freund-Rubin term, as well as other components of the 4-form
field strength, also appear in the generalised vielbein postulates (GVPs)
\cite{deWit:1986mz, Godazgar:2013dma}, and more specifically, 
in the generalised SU(8) connection coefficients $\cQ_{m}{}^{A}{}_{B}$ 
and the generalised non-metricity $\cP_{m \; ABCD}$, we obtain (in our view
the rather elegant) formula (\ref{eq:conjecture-intro}) for $\fr$ that is sextic in the matrix elements of
$\hat\cV$, see Eqn.~\eqref{eq:f-explicit}, by a particular
projection of the internal GVP using components of the 56-bein.

Another projection of the internal GVP gives an ansatz for the
internal components of the field strength. When projecting out
$F_{mnpq}$ from the generalised non-metricity $\mathcal{P}_{m\,ABCD}$,
components of the generalised Christoffel connection
$\mathbf{\Gamma}^\mathcal{P}_{m \,\mathcal{N}}$ contribute, see
Eqn.~\eqref{eq:Fmnpq}. In fact these terms, which correspond to
ambiguities in the language of Ref.~\cite{NP}, {\it remove all terms
  in $F_{mnpq}$ that are not fully antisymmetric} so that $F_{mnpq} =
F_{[mnpq]}$, as required by its compatibility with
(\ref{11dfieldcon}).
Note that, when projecting out the Freund-Rubin term $\fr$ from the
generalised vielbein postulate, components of the generalised
Christoffel connection drop out. In this way we are finally able to
resolve an issue that was left unfinished in Ref.~\cite{deWit:1986iy}:
it is also observed there that one can project out the 4-form field
strength. However, the resulting SU(8) invariant expression, apart
from the ambiguities pointed out in \cite{NP}, turns out to be
unmanageably complicated due to the fact that only part of the
generalised vielbein was known. Nevertheless we can now confirm that
this strategy is correct, and does yield non-linear ans\"atze for the
field strengths of the form fields. In particular, these new ans\"atze
can be more suitable than using the ones for the form fields
themselves.

Furthermore, in section \ref{sec:othercom}, we use the external GVP
and the fermion supersymmetry transformations to find ans\"atze for
the remaining components of the field strength. In particular, we find
new direct and simple ans\"atze for the $F_{\mu \nu \rho m}$ and
$F_{\mu \nu mn}$ components, Eqns.~\eqref{eq:F31ansatzs7} and
\eqref{eq:F22s7}, respectively. We verify the ansatz for $F_{\mu \nu
  \rho m}$ for the SO$(7)^+$ sector.

\section{Non-linear ansatz for the Freund-Rubin term}
\label{sec:non-linear-ansatz}

\subsection{The 56-bein $\mathcal{V}$}
\label{sec:56-bein-mathcalv}

The internal components of the $D=11$ fields are packaged into a
single coset element of E$_{7(7)}/$SU(8), a 56-bein
$(\mathcal{V}_\mathcal{M}{}^{AB}, \cV_{\cM\,AB})$. Here, the
E$_{7(7)}$ index $\cM$ decomposes under GL(7) as
\begin{equation}
  {\cal V}_{\cal M} = \left( {\cal V}_{m8}, {\cal V}_{mn}, {\cal V}^{mn},
    {\cal V}^{m8} \right),
\end{equation}
with GL(7) indices $m,n, \dots= 1, \dots,7$. The components of ${\cal V}$ in
terms of the $D=11$ fields are~\cite{Godazgar:2013dma},
\begin{align}
  \mathcal{V}^{m8}{}_{AB} =& - \frac{\sqrt{2}}{8} \Delta^{-1/2}
  \Gamma^m_{AB}, \label{V11d1} \\
  \mathcal{V}_{mn\;AB} =& - \frac{\sqrt{2}}{8} \Delta^{-1/2} \left(
    \Gamma_{mn\;AB} + 6\sqrt{2} A_{mnp} \Gamma^p_{AB}
  \right),\\
  \mathcal{V}^{mn}{}_{AB} =& - \frac{\sqrt{2}}{8} \cdot \frac{1}{5!}
  \mathring \eta^{mnp_1\cdots p_5} \Delta^{-1/2} \Bigg[
  \Gamma_{p_1\cdots p_5\;AB} + 60\sqrt{2} A_{p_1p_2p_3} \Gamma_{p_4
    p_5\;AB} \nonumber\\ &\qquad\qquad\qquad\qquad\qquad\qquad -
  6!\sqrt{2} \left( A_{qp_1\cdots p_5} - \frac{\sqrt{2}}{4}
    A_{qp_1p_2} A_{p_3p_4p_5} \right) \Gamma^q_{AB}
  \Bigg],\\
  \mathcal{V}_{m8\;AB} =& - \frac{\sqrt{2}}{8} \cdot \frac{1}{7!}
  \mathring \eta^{p_1 \cdots p_7} \Delta^{-1/2} \Bigg[
  (\Gamma_{p_1\cdots p_7}\Gamma_m)_{AB} + 126\sqrt{2}
  A_{mp_1p_2}\Gamma_{p_3\cdots p_7\;AB} \nonumber\\
  &\qquad\qquad\qquad\qquad\qquad\qquad + 3\sqrt{2}\cdot 7! \left(
    A_{mp_1\cdots p_5} + \frac{\sqrt{2}}{4} A_{mp_1p_2}A_{p_3p_4p_5}
  \right) \Gamma_{p_6p_7\;AB} \nonumber\\
  &\qquad\qquad\qquad\qquad\qquad\qquad + \frac{9!}{2} \left(
    A_{mp_1\cdots p_5} + \frac{\sqrt{2}}{12}A_{mp_1p_2}A_{p_3p_4p_5}
  \right) A_{p_6p_7q} \Gamma^q_{AB} \Bigg]. \label{V11d4}
\end{align}
Here, $\Gamma^{a_1\cdots a_n} = \Gamma^{[a_1} \ldots \Gamma^{a_n]}$
are seven-dimensional $8\times8$ $\Gamma$-matrices and
$\Gamma^{m_1\cdots m_n}$ are their curved versions, e.g. $\Gamma^m =
e^m{}_a \Gamma^a$. $A_{pqr}$ and $A_{m_1\cdots m_6}$ are 3-form and
6-form fields, respectively. The $\cV$ given above is an E$_{7(7)}$
matrix because it corresponds to the exponentiation of E$_{7(7)}$ Lie
algebra elements~\cite{Hillmann:2009ci}.

The index $\cM$ that denotes the $\mathbf{56}$ of E$_{7(7)}$ is raised
and lowered with the symplectic metric
$\Omega^{\mathcal{M}\mathcal{N}}$ and its inverse, namely
$$\mathcal{V}^\mathcal{M} =\Omega^{\mathcal{M}\mathcal{N}}
\mathcal{V}_\mathcal{N}.$$
The non-vanishing components of $\Omega^{\mathcal{M}\mathcal{N}}$ are
\begin{equation} 
  \Omega^{mn}{}_{pq} = - \,\Omega_{pq}{}^{mn} = \delta^{mn}_{pq},\qquad 
  \Omega^{m8}{}_{p8} = - \, \Omega_{p8}{}^{m8} =
  \delta^{m8}_{p\, 8} = \frac12 \delta^m_p
\end{equation}
and its inverse is defined by $$ \Omega^{\mathcal{M}\mathcal{P}}
\Omega_{\mathcal{N}\mathcal{P}} = \delta^{\mathcal{M}}_\mathcal{N}.$$
$(\cV_{\cM}{}^{A B} \; \cV_{\cM}{}_{A B})$ is an Sp$(56, \mathbb{R})$ matrix and hence
\begin{equation}
  \label{eq:Vnorm}
  \mathcal{V}^{\mathcal{M}\;AB} \mathcal{V}_{\mathcal{M}\,CD} = i\,
  \delta^{AB}_{CD},\qquad \mathcal{V}^{\mathcal{M}\;AB}
  \mathcal{V}_{\mathcal{M}}{}^{CD} = 0.
\end{equation}
`Curved' SU(8) indices $A,B, \dots$ are raised and lowered by complex
conjugation,
\begin{equation}
  \mathcal{V}_\mathcal{M}{}^{AB} = (\mathcal{V}_{\mathcal{M}\;AB})^*,
  \qquad \mathcal{V}^{\mathcal{M}\;AB} =
  \left(\mathcal{V}^\mathcal{M}{}_{AB}\right)^*,
\end{equation}
while the position of the E$_{7(7)}$ index on $\mathcal{V}$ is not
affected.

The $D=11$ 56-bein is related via the linear ansatz \eqref{eq:scalans}
\cite{Godazgar:2013dma} to the E$_{7(7)}$ matrix that encodes the
scalars of $N=8$ supergravity
\begin{equation}
  \hat{\mathcal{V}} =
  \begin{pmatrix}
    u_{ij}{}^{IJ} & v_{ij\;IJ} \\[1mm] v^{ij\;IJ} & u^{ij}{}_{IJ}
  \end{pmatrix}.
\end{equation}
The 70 scalars and pseudoscalars parametrise $u_{ij}{}^{IJ}(x)$ and
$v_{ij\;IJ}(x)$. In the form above, the 56-bein is given in an SU(8)
basis. However, it turns out to be more convenient to have the 56-bein
such that its E$_{7(7)}$ index is decomposed in an SL(8)
basis:~\footnote{See, for example, Ref.~\cite{Bossard:2010dq} for more
  explanation.}
\begin{equation} \label{eq:cVhat} \hat{\mathcal{V}}_{\cM}{}^{ij} =
  \frac{1}{\sqrt{2}} 
  \begin{pmatrix} u^{ij}{}_{IJ} - v^{ijIJ} \\[2mm]
    -i (u^{ij}{}_{IJ} + v^{ijIJ}) 
  \end{pmatrix}\; , \quad \hat{\mathcal{V}}_{\cM\, ij} \equiv \big(
  \hat{\mathcal{V}}_{\cM}{}^{ij}\big)^* = \frac{1}{\sqrt{2}}
  \begin{pmatrix}
    u_{ij}{}^{IJ} - v_{ijIJ} \\[2mm]
    i (u_{ij}{}^{IJ} + v_{ijIJ})
  \end{pmatrix}.
\end{equation}

In relating the $d=4$ 56-bein to the eleven-dimensional one given above,  
one must in principle take into account a compensating SU(8) rotation
depending on all eleven coordinates, as explained
in Ref.~\cite{deWit:1986mz}. However, in the remainder we will deal only
with quantities where the SU(8) indices are fully contracted, and this
SU(8) rotation drops out. Keeping this in mind, the explicit
dependence of the components on the $d=4$ fields is
\cite{Godazgar:2013pfa}
\begin{gather}
  \label{eq:4d-vielbeinB}
  \mathcal{V}^{m8}{}_{ij}(x,y) = \frac{\sqrt{2}i}{8} K^{m\;IJ}(y)
  \left( u_{ij}{}^{IJ} + v_{ij\;IJ}
  \right)(x),\\ \label{eq:4d-vielbeinC} \mathcal{V}_{mn\;ij}(x,y)=
  -\frac{\sqrt{2}}{8} K_{mn}{}^{IJ}(y) \left( u_{ij}{}^{IJ} -
    v_{ij\;IJ}
  \right)(x),\\
  \mathcal{V}^{mn}{}_{ij} (x,y)= \frac{\sqrt{2}i}{8\cdot5!} \mathring
  \eta^{mn p_1\cdots p_5} \left( K_{p_1\cdots p_5}{}^{IJ} - 6\cdot
    6!\, \mathring \zeta_{p_1\cdots p_5 q} K^{q\; IJ} \right)(y)
  \left( u_{ij}{}^{IJ} + v_{ij\;IJ}
  \right)(x),\\
  \mathcal{V}_{m8\;ij} (x,y) = \frac{\sqrt{2}}{8} \left( K_m{}^{IJ} +
    6 \mathring \eta^{p_1\cdots p_7} \mathring \zeta_{p_1\cdots p_6}
    K_{p_7m}{}^{IJ} \right)(y) \left( u_{ij}{}^{IJ} - v_{ij\;IJ}
  \right)(x),
  \label{eq:4d-vielbeinE}
\end{gather}
where $K_m^{IJ}(y)$ are the Killing vectors on the round seven-sphere,
\begin{align}
  K_m{}^{IJ}& = i \bar{\eta}^I \Gamma_m \eta^J, \quad K_{mn}{}^{IJ} =
  -\frac{1}{m_7} \mathring D_m K_n{}^{IJ} = \bar{\eta}^I \Gamma_{mn}
  \eta^J, \label{eq:Kid} \\[1mm]
  & K_{m_1\cdots m_5}{}^{IJ} = i \bar{\eta}^I \Gamma_{m_1\cdots m_5} \eta^J =
  -\frac{1}{2} \mathring \eta_{m_1\cdots m_7} K^{m_6 m_7\;IJ}.
\end{align}
The derivative operator $\Do_{m}$ is the covariant derivative with
respect to the Christoffel connection on the round sphere and
$\eta^{I}$ are the eight Killing spinors on $S^7$. Additionally,
$\mathring\zeta_{m_1\cdots m_6}$ is implicitly defined by
\begin{equation}
  7! \mathring D_{[m_1} \mathring\zeta_{m_2\cdots m_7]} = m_7\, \mathring
  \eta_{m_1\cdots m_7}.
\end{equation}
Furthermore the normalisations in \eqref{eq:4d-vielbeinB} --
\eqref{eq:4d-vielbeinE} have been chosen so that this vielbein is
indeed normalised according to Eqn.~(\ref{eq:Vnorm}). These
expressions are sufficient to derive all non-linear ans\"atze.

\subsection{Generalised vielbein postulate}
\label{sec:gener-vielb-post}

The generalised vielbein postulates are differential constraints on the 56-bein in terms of generalised connections including an SU(8) connection, a generalised E$_{7(7)}$ connection and a generalised non-metricity. Using the GL(7) decomposition of the 56-bein, \eqref{V11d4}, its derivative can be grouped into objects that satisfy the correct transformation properties, namely the generalised connections in Refs.~\cite{deWit:1986mz, Godazgar:2013dma}. The crucial feature of the generalised connections that we utilise in order to derive our ans\"atze is that they are parametrised by components of the 4-form field strength. 
This is a somewhat different approach to the deductive
approach of Ref.~\cite{Coimbra:2011ky}. There, the generalised
connections are found by requiring a torsion-free compatible
connection (in contrast to usual differential geometry, this does not
uniquely specify the connections \cite{Coimbra:2011ky}). The
generalised connections in Ref.~\cite{Coimbra:2011ky} are
nevertheless related \cite{Godazgar:2014nqa} to the generalised
connections in Ref.~\cite{Godazgar:2013dma}; as are the connections in
exceptional field theory \cite{Godazgar:2014nqa}, where the emphasis
is on connections that are expressed in terms of the 56-bein of
exceptional field theory \cite{Hohm:2013uia}.

A distinctive feature of the generalised connections that we use is that they are valued along the first seven directions in a GL(7) decomposition, as is clear from Eqn.~\eqref{eq:GVP}. Note that this is not a consequence of the derivative index running over seven directions, but rather a consequence of working with a generalised non-metricity rather than torsion-free compatible
connections \cite{Godazgar:2014nqa}, which are valued in the {\textbf{56}} even when the base space is not extended as in Ref.~\cite{Coimbra:2011ky}. However, for us it is precisely the SU(8) covariant generalised non-metricity that yields the new non-linear ans\"atze.    

The 56-bein $\mathcal{V}_\mathcal{M}$ satisfies, in particular, the
internal GVP \cite{Godazgar:2013dma,
  Godazgar:2014sla}
\begin{equation}
  \label{eq:GVP}
  \partial_m \mathcal{V}_{\mathcal{N}\,AB} -
  \mathbf{\Gamma}_{m \, \mathcal{N}}{}^\mathcal{P}
  \mathcal{V}_{\mathcal{P}\,AB} + \mathcal{Q}^C_{m\,[A}
  \mathcal{V}_{\mathcal{N}\,B]C} \,=\, 
  \mathcal{P}_{m\,ABCD} \mathcal{V}_{\mathcal{N}}{}^{CD},
\end{equation}
where $\mathcal{Q}^A_{m\,B}$ is the generalised SU(8) connection. The SU(8) tensor
$\mathcal{P}_{m\,ABCD}$ is the `generalised non-metricity', which `measures' the
failure of the metric 
$$
\cG_{\cM\cN} \,\equiv\, \cV_\cM{}^{AB}  \cV_{\cN\,AB}  + \cV_\cN{}^{AB}  \cV_{\cM\,AB}
$$ 
to be covariantly constant under the generalised covariant derivative.~\footnote{As
  explained in Ref.~\cite{Godazgar:2014nqa} the non-metricity can be absorbed into
  the connections, at the price of introducing components $\cQ^A_{\cM \,B}$ and
  $\mathbf{\Gamma}_{\cM\cN}^\cP$ along directions $\cM\neq m$.}
$\mathbf{\Gamma}_{m \mathcal{N}}{}^\mathcal{P}$ is the E$_{7(7)}$
generalised Christoffel connection with components
\begin{gather}
  (\mathbf{\Gamma}_m)_{p8}{}^{q8} = - (\mathbf{\Gamma}_m)^{q8}{}_{p8}
  = \frac{1}{2} \Gamma^q_{mp} + \frac{1}{4} \Gamma^n_{mn} \delta^q_p,
  \qquad (\mathbf{\Gamma}_m)_{pq}{}^{rs} =
  -(\mathbf{\Gamma}_m)^{rs}{}_{pq} = 2\Gamma_{m[p}^{[r}
  \delta^{s]}_{q]} - \frac{1}{2} \Gamma^n_{mn}
  \delta^{rs}_{pq},\nonumber\\[2mm]
  (\mathbf{\Gamma}_m)_{p8}{}^{rs} = - (\mathbf{\Gamma}_m)^{rs}{}_{p8}
  = 3\sqrt{2} \mathring \eta^{rst_1\cdots t_5} \Xi_{m|p t_1\cdots
    t_5}, \nonumber\\[2mm]
  (\mathbf{\Gamma}_m)_{pqr8} = (\mathbf{\Gamma}_m)_{r8pq} = 3\sqrt{2}
  \Xi_{m|pqr}, \qquad (\mathbf{\Gamma}_m)^{pqrs} = \frac{1}{\sqrt{2}}
  \mathring \eta^{pqrs t_1 t_2 t_3} \Xi_{m| t_1 t_2 t_3}.
  \label{eq:E7-Christoffels}
\end{gather}
Here,
$$
\Gamma^p_{mn}(x,y)\equiv \frac12 g^{pq}(\partial_{m} g_{nq}
+ \partial_{n} g_{mq} - \partial_q g_{mn})
$$
denotes the usual Christoffel connection defined with respect to the
metric $g_{mn}(x,y)$. The quantities $\Xi_{m|pqr}(x,y)$ and
$\Xi_{m|p_1\cdots p_6}(x,y)$ are \cite{Godazgar:2014sla}
\begin{align}
  &\Xi_{m|pqr} = D_m A_{pqr} - \frac{1}{4!} F_{mpqr}, \\
  &\Xi_{m|p_1\cdots p_6} = D_m A_{p_1\cdots p_6} + \frac{\sqrt{2}}{48}
  F_{m[p_1 p_2 p_3} A_{p_4 p_5 p_6]} \notag \\[2mm]
  & \hspace{45mm} - \frac{\sqrt{2}}{2} \left( D_m A_{[p_1 p_2 p_3} -
    \frac{1}{4!}F_{m[p_1 p_2 p_3} \right) A_{p_4 p_5 p_6]} -
  \frac{1}{7!} F_{m p_1 \cdots p_6},
\end{align}
where $D_m$ denotes the covariant derivative with respect to the
Christoffel connection $\Gamma_{mn}^p$. From the definitions above it
is clear that
$$
\Xi_{[m|npq]} = \Xi_{[m|p_1\cdots p_6]} = 0.
$$
We also note that under generalised diffeomorphisms (including the
two- and five-form gauge transformations) all connection coefficients
transform with second derivatives, just like the standard Christoffel
connection.

In a non-trivial background (such as the compactification on $S^7$),
all E$_{7(7)}$ Christoffel connections decompose into a background
connection
$\mathring{\mathbf{\Gamma}}^\mathcal{P}_{m \, \mathcal{N}}$ and
a variation
$\hat{\mathbf{\Gamma}}^\mathcal{P}_{m\, \mathcal{N}}$,
\begin{equation}\label{Gamma0}
  \mathbf \Gamma_{m \, \mathcal{N}}{}^\mathcal{P} \,=\,
  \mathring{\mathbf{\Gamma}}_{m\, \mathcal{N}}{}^\mathcal{P} \,+\,
  \hat {\mathbf{\Gamma}}_{m \,\mathcal{N}}{}^\mathcal{P}.
\end{equation}
For the $S^7$ compactification we will see, Eqns.~\eqref{eq:ChristoffelBackground1}--\eqref{eq:ChristoffelBackground3},
that the background connection is not only given by the standard
covariantisation with respect to $\mathring\Gamma^p_{mn}$, but that it also requires a non-vanishing component $\mathring \Xi_{m|p_1\cdots
  p_6}$.

The generalised spin connection $\mathcal{Q}^A_{m\,B}$ and
non-metricity $\mathcal{P}_{m\;ABCD}$ are expressed in terms of the
$D=11$ fields as follows \cite{deWit:1986mz}:
\begin{gather}
  \label{eq:Q}
  \mathcal{Q}^A_{m\,B} \,=\, -\frac{1}{2} \omega_{m\,ab}
  \Gamma^{ab}_{AB} \,+\, \frac{\sqrt{2}}{14} i \Delta^2 \,
  \fr \Gamma_{m\;AB} \,-\,
  \frac{\sqrt{2}}{48} F_{mnpq} \Gamma^{npq}_{AB},\\[2mm]
  \label{eq:P}
  \mathcal{P}_{m\,ABCD} \,=\, \frac{\sqrt{2}}{56} i
  \Delta^2 \, \fr \Gamma_{mn[AB}\Gamma^n_{CD]} \, + \,
  \frac{\sqrt{2}}{32} F_{mnpq} \Gamma^n_{[AB} \Gamma^{pq}_{CD]},
\end{gather}
where $\omega_{m\;ab}$ is the SO(7) spin-connection. The internal GVP,
(\ref{eq:GVP}), provides a non-linear ansatz for $\fr$, given that
$\cP_{m}$ depends on $\fr$.  From
Eqn.~(\ref{eq:GVP}), we find
\begin{equation} \label{eq:Pm} \mathcal{P}_{m\;ABCD} \, = \,
  -i\cV^\cN{}_{CD} {\mathbf{D}}_m \cV_{\cN \, AB} \,\equiv \, -i
  \mathcal{V}^\mathcal{N}{}_{CD} \partial_m
  \mathcal{V}_{\mathcal{N}\;AB} + i
  \mathbf{\Gamma}_{m \,\mathcal{N}}{}^\mathcal{P}
  \mathcal{V}^\mathcal{N}{}_{CD} \mathcal{V}_{\mathcal{P}\;AB}
\end{equation}
and project out the Freund-Rubin term using the $D=11$ vielbein
components,
\begin{equation}
  \fr = -\frac{8\sqrt{2}i}{3} \mathcal{V}^{m8\;EF}
  \mathcal{V}^{p8}{}_{EF} \mathcal{V}_{pq}{}^{AB} \mathcal{V}^{q8\;CD}
  \mathcal{P}_{m\,ABCD}.
\end{equation}
Note that in Eqn.~(\ref{eq:Pm}), we defined the full covariant
derivative $\mathbf{D}_m$ with respect to the full E$_{7(7)}$ Christoffel
connection. We denote the covariant derivative associated with the
full background connection
$\mathring{\mathbf{\Gamma}}^\mathcal{P}_{m \,\mathcal{N}}$, 
$\mathring{\mathbf{D}}_m$.

Substituting the expression for $\cP_{m}$ from Eqn.~\eqref{eq:Pm},
this projection has the following convenient property: as a result of
contracting out all SU(8) indices all the generalised connection 
components (\ref{eq:E7-Christoffels}) drop out in $\fr$.
For this reason we can use any connection; we choose to work
with the background connection for convenience. Note that this
is {\it not true for other projections, in particular the 4-form field
  strength $F_{mnpq}$}. In section \ref{sec:outlook}, we give a new
ansatz for $F_{mnpq}$ that takes these ``ambiguities'' into account. Thus,
\begin{equation}
  \label{eq:f11d}
  \fr = -\frac{8\sqrt{2}}{3} \mathcal{V}^{m8\;EF}
  \mathcal{V}^{p8}{}_{EF} \mathcal{V}_{pq}{}^{AB} \mathcal{V}^{q8\;CD}
  \mathcal{V}^\mathcal{N}{}_{CD} \mathring{\mathbf{D}}_m
  \mathcal{V}_{\mathcal{N}\;AB}.
\end{equation}

\subsection{The Freund-Rubin term in terms of $d=4$ fields}
\label{sec:freund-rubin-factor-1}

We convert curved SU(8) indices $A,B, \dots$ into flat SU(8) indices
$i,j, \dots$ (\emph{cf.}~Eqn.~\eqref{eq:scalans}) by means of the orthonormal Killing spinors on the round
sphere $\eta^i_A$,
\begin{gather}
  \label{eq:P_ijkl}
  \mathcal{P}_{m\;ijkl} = -i \, \mathcal{V}^\mathcal{N}{}_{kl}
  \mathring{\mathbf{D}}_m \mathcal{V}_{\mathcal{N}\;ij} + i \,
  \mathbf{\hat{\Gamma}}_{m \, \mathcal{N}}{}^\mathcal{P}
  \mathcal{V}^\mathcal{N}{}_{kl} \mathcal{V}_{\mathcal{P}\;ij},\\
  \label{eq:f}
  \fr = -\frac{8\sqrt{2}}{3} \mathcal{V}^{m8\;rs}
  \mathcal{V}^{p8}{}_{rs} \mathcal{V}_{pq}{}^{ij} \mathcal{V}^{q8\;kl}
  \mathcal{V}^\mathcal{N}{}_{kl} \mathring{\mathbf{D}}_m
  \mathcal{V}_{\mathcal{N}\;ij}.
\end{gather}
Here, we used the split \eqref{Gamma0} for the $S^7$ background, with
the only non-vanishing Christoffel connection components
\begin{eqnarray} \label{eq:ChristoffelBackground1}
  (\mathring{\mathbf{\Gamma}}_m)_{p8}{}^{q8} &=& -
  (\mathring{\mathbf{\Gamma}}_m)^{q8}{}_{p8} \,=\, \frac{1}{2}
  \mathring\Gamma^q_{mp} + \frac{1}{4} \mathring\Gamma^n_{mn}
  \delta^q_p, \\[2mm] \label{eq:ChristoffelBackground2}
  (\mathring{\mathbf{\Gamma}}_m)_{pq}{}^{rs} &=&
  -(\mathring{\mathbf{\Gamma}}_m)^{rs}{}_{pq} \,=\,
  2\mathring\Gamma_{m[p}^{[r} \delta^{s]}_{q]} - \frac{1}{2}
  \mathring\Gamma^n_{mn} \delta^{rs}_{pq}
\end{eqnarray}
and
\begin{equation} \label{eq:ChristoffelBackground3} (\mathbf{\mathring
    \Gamma}_m)_{p8}{}^{rs} = -
  (\mathbf{\mathring\Gamma}_m)^{rs}{}_{p8} = 3\sqrt{2} \eta^{rs t_1
    \cdots t_5} \mathring \Xi_{m|p t_1\cdots t_5},
\end{equation}
with
\begin{equation}
  \mathring \Xi_{m|n_1\cdots n_6} \, \equiv \,
  \mathring D_m \mathring
  A_{n_1\cdots n_6} - \mathring D_{[m} \mathring A_{n_1\cdots n_6]} =
  3\sqrt{2} \left( \Do_{[m} \zetao_{ n_1\cdots n_6]} - \mathring D_m \mathring \zeta_{n_1\cdots n_6} \right).
\end{equation}

Thus, the evaluation of the Freund-Rubin term requires an evaluation of
the Maurer-Cartan form of the 56-bein. This can simply be calculated
using Eqns.~\eqref{eq:4d-vielbeinB}--\eqref{eq:4d-vielbeinE},
\begin{align} 
  \mathcal{V}^\mathcal{N}{}_{kl} \mathring D_m
  \mathcal{V}_{\mathcal{N}\;ij} =& \frac{3}{28} i m_7 K_{mn}{}^{[IJ}
  K^{n\,KL]} \left( u_{ij}{}^{IJ} u_{kl}{}^{KL} - v_{ij\,IJ}
    v_{kl\,KL}
  \right) \notag \\[2mm]
  & \hspace{5mm} + \frac{4}{7} i m_7 K_{m}{}^{IJ} \left( v_{ij\,MJ}
    u_{kl}{}^{MI} - u_{ij}{}^{MJ} v_{kl\,MI} \right) \nonumber\\[2mm]&
  \hspace{10mm} - 12\, \etao^{n_1 \dots n_7} \left( \Do_{m}
    \zetao_{n_1 \dots n_6} - \Do_{[m} \zetao_{n_1 \dots n_6]} \right)
  \left( \cV^{p8}{}_{ij} \cV_{n_7 p \, kl} + \cV^{p8}{}_{kl} \cV_{n_7
      p \, ij} \right),
  \label{eq:vdvsoln}
\end{align}
where $\Do_m$ is the usual $S^7$ covariant derivative. The last term
on the right-hand side of the above expression exactly cancels the
contribution of the generalised connection term coming from
$\mathring\Xi_{m|p_1\cdots p_6}$ evaluated in
$\mathcal{V}^\mathcal{N}{}_{kl} \mathring D_m
\mathcal{V}_{\mathcal{N}\;ij}$. Namely, at the background value of the
fields, where $\hat{\mathbf{\Gamma}}^\mathcal{P}_{m \, \mathcal{N}} = 0$, $\cP_{m \; ijkl}$ given by Eqn.~\eqref{eq:P_ijkl}
is equal to the first two terms in Eqn.~\eqref{eq:vdvsoln},
reproducing the solution given in equation (3.19) of
Ref.~\cite{deWit:1986iy}.~\footnote{In Ref.~\cite{deWit:1986iy},
  $\cP_{m \; ijkl}$ is denoted by ${\mathcal{A}}_{m \; ijkl}.$}
Otherwise, away from the SO(8) invariant vacuum, the solution is
modified by the generalised connection terms
$\hat{\mathbf{\Gamma}}^\mathcal{P}_{m \, \mathcal{N}}$. These
are the ``ambiguities'' that leave the supersymmetry transformations
unchanged \cite{NP}. Therefore, the solution proposed in
Ref.~\cite{deWit:1986iy} is consistent with the supersymmetry
transformations but does not reproduce the field strength components
$F_{mnpq}$. In generalised geometry, this is manifested in the lack of
a unique torsion-free, metric-compatible generalised connection
\cite{Coimbra:2011ky}; see also Ref.~\cite{Godazgar:2014nqa} where
this relation was explored.

In fact, equation \eqref{eq:vdvsoln} points to the necessity of using
a background connection that accounts for the fact that the
Freund-Rubin parameter is non-zero at the background. This background
connection includes generalised connection components such that
\begin{align} 
  \mathcal{V}^\mathcal{N}{}_{kl} \mathring{\mathbf{D}}_m
  \mathcal{V}_{\mathcal{N}\;ij} =& \frac{3}{28} i m_7 K_{mn}{}^{[IJ}
  K^{n\,KL]} \left( u_{ij}{}^{IJ} u_{kl}{}^{KL} - v_{ij\,IJ}
    v_{kl\,KL}
  \right) \notag \\[2mm]
  & \hspace{5mm} + \frac{4}{7} i m_7 K_{m}{}^{IJ} \left( v_{ij\,MJ}
    u_{kl}{}^{MI} - u_{ij}{}^{MJ} v_{kl\,MI} \right).
  \label{eq:vdvsoln2}
\end{align}
However, since our identities, e.g.\ \eqref{eq:Kid}, are written in
terms of the usual $S^7$ covariant derivative $\mathring D_m$, 
we use this connection for convenience.

From Eqns.~\eqref{eq:P_ijkl} and \eqref{eq:vdvsoln2}, we can now see
exactly how the solution given in equation (3.19) of Ref.~\cite{deWit:1986iy} for $\cP_{m \; ijkl}$ is modified by the
generalised connection coefficients. It is clear from
Eqn.~\eqref{eq:P_ijkl} that the role of the generalised connection
term is to fully antisymmetrise $\Do_{m} A_{npq}$ and $\Do_{m_1}
A_{m_2 \dots m_7}$ terms coming from $\Do_{m} \cV_{\cM}$. This gives the
field strength components $F_{mnpq}$ and $F_{m_1 \dots m_7 }$ in
$\cP_{m \; ijkl}$---without the generalised vielbein postulate this
task would be an unwieldy problem.

We now make use of Eqn.~\eqref{eq:vdvsoln}, remembering that the contributions from the generalised connections vanish, and insert the explicit
formulae for the vielbein components,
(\ref{eq:4d-vielbeinB})--(\ref{eq:4d-vielbeinE}), into the expression
for the Freund-Rubin term, (\ref{eq:f}). Defining
\begin{align}
  \label{eq:X}
  \rX_{rs}{}^{ijkl}(x,y) &= K^{IJKL}(y) \left( u_{rs}{}^{IM} +
    v_{rs\;IM} \right) \left( u^{ij}{}_{[JK} u^{kl}{}_{LM]} -
    v^{ij[JK} v^{kl\;LM]}
  \right)(x) ,\\
  \label{eq:Y}
  \rY_{rs}{}^{ijkl}(x,y) &= K^{m \,IJ}K_{m}{}^{KL}(y) \left(
    u_{rs}{}^{IJ} + v_{rs\;IJ} \right) \left( u^{ij}{}_{KM} v^{kl\;LM}
    - v^{ij\;KM} u^{kl}{}_{LM} \right)(x) ,
\end{align}
where  $K^{IJKL}(y) = K_m{}^{[IJ}(y) K^{m\;KL]}(y)$, we find that
\begin{align}
  \mathcal{V}^{m8\;rs} \mathcal{V}^\mathcal{N}{}_{kl} \mathring{{D}}_m
  \mathcal{V}_{\mathcal{N}\;ij} &= - \frac{\sqrt{2} m_7}{28} \left( 3
    \rX^{rs}{}_{ijkl} - 2 \rY^{rs}{}_{ijkl}
  \right), \\
  \mathcal{V}^{p8}{}_{rs} \mathcal{V}_{pq}{}^{[ij}
  \mathcal{V}^{q8\;kl]} &= \frac{\sqrt{2}}{64} \left( 2
    \rX_{rs}{}^{ijkl} + \rY_{rs}{}^{ijkl} \right).
\end{align}
Thus, the Freund-Rubin term is
\begin{align}
  \fr(x,y) = \frac{m_7}{168\sqrt{2}} \Bigl( 3 \rX^{rs}{}_{ijkl} - 2
  \rY^{rs}{}_{ijkl} \Bigr) \Bigl( 2 \rX_{rs}{}^{ijkl} +
  \rY_{rs}{}^{ijkl} \Bigr)(x,y).
  \label{eq:f-explicit}
\end{align}

\section{Examples}
\label{sec:testing-non-linear}

In the following, we evaluate the Freund-Rubin term
(\ref{eq:f-explicit}) for the G$_2$ invariant sector \cite{Warner:1983vz, Warner:1983du}. We refer the reader to appendices \ref{app:freund-rubin-so3t} and \ref{app:freund-rubin-su4} for the  Freund-Rubin term for the  SO(3)$\times$SO(3) and SU(4)$^-$ invariant sectors. At stationary
points, $\fr$ is proportional to the scalar potential. This has
already been noted in Ref.~\cite{NP}. Eqn.~\eqref{eq:f-explicit} now
gives a general expression for $\fr$ away from stationary points. In
the following examples, we will see that this expression always
consists of two parts: the first part is proportional to
the scalar potential $V$  --  this has been verified for many stationary
points \cite{NP}. The second part is proportional to a variation of
the potential and depends on internal coordinates. Thus, the
Freund-Rubin term is only constant at stationary points. In uplifts of
flows the Freund-Rubin term will, in general, be both $x$ and
$y$-dependent.

\subsection{Freund-Rubin term in the G$_2$ invariant sector}
\label{sec:freund-rubin-factor}

In a `unitary gauge,' the 56-bein takes the special form
\begin{equation}
  \label{G2vielbein}
  \mathcal{V} =
  \begin{pmatrix}
    u_{IJ}{}^{KL} & v_{IJ\;KL} \\[1mm] v^{IJ\;KL} & u^{IJ}{}_{KL}
  \end{pmatrix} = \exp
  \begin{pmatrix}
    0 & \phi_{IJKL} \\ \phi^{IJKL} & 0
  \end{pmatrix}.
\end{equation}
For the G$_2$ invariant sector
\begin{equation}
  \phi_{IJKL} \,\equiv \, \phi_{IJKL}(\alpha,\lambda) \,=\, \frac{\lambda}{2} \left(
  \cos \alpha\, C_+^{IJKL}  \,+\, i \sin\alpha\, C_-^{IJKL} \right)
\end{equation}
with the SO(7)$^+$ and SO(7)$^-$ invariant 4-form tensors
$C_+^{IJKL}$ and $C_-^{IJKL}$, respectively. The common invariance
group is G$_2$ = SO(7)$^+ \cap$ SO(7)$^-$. 

The scalar potential for the G$_2$ invariant sector, calculated from Eqn.~\eqref{eq:potential}, reads
\begin{align}
  \label{eq:V-G2}
  V(\alpha,\lambda) =&\; 2g^2\left[(7v^4 - 7v^2 + 3)c^3s^4 + (4v^2 -
    7)v^5s^7 + c^5s^2 + 7v^3c^2s^5 - 3c^3\right] \nonumber\\[2mm]
  =&\; 2(c + vs)^2 \left(
    7v^3s^3 + 4v^5s^5 - 14cv^2s^2 - 8cv^4s^4 + 14c^2vs \right. \notag \\[2mm]
  & \hspace{60mm} \left. + 5c^2v^3s^3 - 7c^3 + 5c^3v^2s^2 - 8c^4vs +
    4c^5 \right).
\end{align}
Here, $g$ is the gauge coupling constant and
\begin{equation}
  c = \cosh2\lambda,\qquad s = \sinh2\lambda,\qquad v = \cos \alpha.
\end{equation}
Taking the derivative of the potential with respect to $\alpha$ and
$\lambda$ yields
\begin{align}
  \label{eq:dVda-G2}
  & \frac{\mathrm{d}V}{\mathrm{d}\alpha} = -14 v s^2 \sin \alpha (c +
  vs) \left( 5v^2s^2 + 4v^4s^4 - 5cvs - 4cv^3s^3 + 2c^2 - c^2 v^2s^2 +
    5c^3 v s - 2c^4 \right),\\[2mm]
  & \frac{\mathrm{d}V}{\mathrm{d}\lambda} = 28\, \frac{c}{s} (c + vs)
  ( 2v^2 s^2 + 7 v^4 s^4 + 4 v^6 s^6 - 5 c vs - 10c v^3 s^3 - 4c
  v^5s^5
  + 5c^2 \nonumber\\[2mm]
  & \hspace{35mm} + c^2 v^2 s^2 - 3 c^2 v^4 s^4 + 9 c^3 v s + 10 c^3
  v^3 s^3 - 9c^4 - 3 c^4 v^2 s^2 - 4 c^5 v s + 4 c^6).
  \label{eq:dVdl-G2}
\end{align}

We write the $ u$ and $v$ tensors in the following basis of G$_2$
invariants~\cite{deWit:1984nz}
\begin{align}
  \label{eq:G2-basis}
  \delta^{IJ}_{KL}, \qquad C_+^{IJKL}, \qquad C_-^{IJKL}, \qquad D_\pm
  = \frac{1}{2} \left( C_+^{IJMN} C_-^{MNKL} \pm C_-^{IJMN} C_+^{MNKL}
  \right).
\end{align}
Here, $C_+^{IJKL}$ is selfdual and $C_-^{IJKL}$ is anti-selfdual.
Having chosen a symmetric gauge for the $d=4$ 56-bein, we do not
distinguish between SU(8) and SO(8) indices. We find
\cite{deWit:1984nz, Godazgar:2013nma}
\begin{align}
  \label{eq:u-G2}
  u_{IJ}{}^{KL}(\lambda,\alpha) =& p^3 \delta^{KL}_{IJ} + \frac{1}{2} p
  q^2 \cos^2\alpha C_+^{IJKL} - \frac{1}{2} p q^2 \sin^2 \alpha
  C_-^{IJKL} - \frac{1}{8} i p q^2 \sin 2\alpha D_-^{IJKL}, \\[2mm]
  \label{eq:v-G2}
  v_{IJKL}(\lambda,\alpha) =& q^3(\cos^3\alpha - i \sin^3 \alpha)
  \delta^{IJ}_{KL} + \frac{1}{2} p^2 q \cos \alpha C_+^{IJKL} +
  \frac{1}{2} i p^2 q \sin \alpha C_-^{IJKL} \notag \\[2mm]
  & \hspace{70mm} - \frac{1}{8} q^3 \sin 2\alpha (\sin \alpha - i \cos
  \alpha) D_+^{IJKL}.
\end{align}
The $x$-dependence is kept in $\lambda= \lambda(x)$ through
\begin{equation}
  p = \cosh \lambda,\qquad
  q = \sinh \lambda.
\end{equation}
$u^{IJ}{}_{KL}$ and $v^{IJKL}$ are obtained from the above equations
by complex conjugation.

Plugging the explicit form of the $u$ and $v$ tensors into the
expression of the Freund-Rubin term and identifying 
\begin{equation}
  \xi (y) = - \frac{1}{16} C_+^{IJKL} K_{m}{}^{IJ} K^{m \, KL},
\end{equation}
we find the Freund-Rubin term in the G$_2$ invariant sector:
\begin{align}
  \label{eq:f-G2-explicit}
  \fr =& -\sqrt{2} m_7 (c + vs)^2 \left( 7 v^3 s^3 + 4 v^5 s^5 - 14 c
    v^2 s^2 - 8 c v^4 s^4 + 14 c^2 vs
  \right. \notag \\[2mm]
  & \hspace{70mm} \left. + 5 c^2 v^3 s^3 - 7 c^3 + 5 c^3 v^2 s^2 - 8
    c^4 vs + 4 c^5 \right) \nonumber\\[1mm]
  &+ \frac{\sqrt2}{3}\, m_7\, \xi\, (c + vs)^2 c v s \left( 3 v s + 2
    v^3 s^3 - 3 c - c v^2 s^2 - c^2 v s + 2 c^3 \right).
\end{align}
While the first two lines are $y$-independent, all the $y$-dependence
here is contained in the factor $\xi(y)$ in the last line. Using
Eqns.~(\ref{eq:V-G2}), (\ref{eq:dVda-G2}) and (\ref{eq:dVdl-G2}), the
above expression can be rewritten as
\begin{equation}
  \label{eq:f-G2}
  \fr = \frac{m_7}{\sqrt{2}g^2} \left( - V + \frac{\xi}{21s}
    \left( \frac{s \cos \alpha}{2}
      \frac{\mathrm{d}V}{\mathrm{d}\lambda} - c \sin \alpha
      \frac{\mathrm{d}V}{\mathrm{d}\alpha}\right) \right).
\end{equation}

This result is exactly of the expected form. The term proportional to
the scalar potential is coordinate invariant. All other terms are
proportional to the derivatives of $V$ with respect to $\alpha$ and
$\lambda$ and thus vanish at the stationary points, that is, when the
equations of motion are obeyed. Off-shell, there is a linear
dependence on $\xi(y)$ so the extra terms do depend on internal
coordinates. Furthermore, $\fr$ is $x$-dependent via $s,c$ and
$\alpha$. Note that the G$_2$ invariant sector also includes as
special cases the SO(7)$^{\pm}$ invariant sectors for appropriate
values of $\alpha$:
\begin{equation}
  \fr= \begin{cases}
    \frac{m_7}{\sqrt{2}g^2} \left( - V + \frac{\xi}{42}
      \frac{\mathrm{d}V}{\mathrm{d}\lambda} \right)\Big|_{v=1} &
    \text{SO(7)}^{+} \\[3mm] 
    - \frac{m_7}{\sqrt{2}g^2}  V\Big|_{v=0}  & \text{SO(7)}^{-}
 \end{cases}. \label{fr:so7}
\end{equation}
(recall that $dV/d\alpha$ vanishes for $v=0$).

We repeat this calculation in appendices \ref{app:freund-rubin-so3t} and \ref{app:freund-rubin-su4} for the
SO(3)$\times$SO(3) and SU(4)$^-$ invariant sectors and find
expressions similar to Eqns.~(\ref{eq:f-G2}) and \eqref{fr:so7}. Motivated by these
results we state a general conjecture for the Freund-Rubin term in section \ref{sec:shell-conj-freund}.

\section{Ans\"atze for other components of the 4-form field strength}
\label{sec:othercom}

Given the new ansatz for the Freund-Rubin term, a natural question
that arises is whether similar ans\"atze for the other components of
the 4-form field strength can also be teased out of the generalised
vielbein postulates. The generalised spin connection and non-metricity
from Eqns.~\eqref{eq:Q} and \eqref{eq:P} in the internal GVP depend on
$F_{mnpq}$ as well as $\fr$. Therefore, one can also project onto the
component giving $F_{mnpq}$. Indeed this is done in
Refs.~\cite{deWit:1986iy},\cite{deWit:1986mz} using only the original
generalised vielbein $e^{m}_{AB}$. However, we use the full 56-bein
and its various components and take account of the generalised
connection term. We can project onto the $F_{mnpq}$ term by performing
the following contraction of $\cP_{m \, ABCD}$ with components of the 56-bein:
\begin{equation}
  \cP_{m \, ABCD} \left(\cV^{r8 \;AB} \cV_{pq}{}^{CD} + \frac{1}{4!} \epsilon^{ABCDEFGH}  \cV^{r8}{}_{EF} \cV_{pq \;GH}\right) = \frac{1}{16} \Delta^{-1} g^{rn} F_{mnpq}.
\end{equation}
Therefore, from Eqn.~\eqref{eq:Pm}, we find that the uplift ansatz for
$F_{mnpq}$ is given by
\begin{equation} \label{eq:Fmnpq} \Delta^{-1} g^{nr} F_{mnpq} = - 16 i
  \left( \mathcal{V}^\mathcal{N}{}_{ij} \partial_m
    \mathcal{V}_{\mathcal{N}\;kl} -
    \mathbf{\Gamma}_{m \,\mathcal{N}}{}^\mathcal{P}
    \mathcal{V}^\mathcal{N}{}_{ij} \mathcal{V}_{\mathcal{P}\;kl}
  \right) \cV^{r8 \;ij} \cV_{pq}{}^{kl} + h.c..
\end{equation}

The ansatz above is not as direct as the formula for the Freund-Rubin
term \eqref{eq:f11d}. Firstly, as with the non-linear flux ansatz
\cite{dWN13} one needs to invert the metric to obtain
$F_{mnpq}$.~\footnote{In fact, contracting $ \cP_{m \, ABCD}$ with
  other components of the 56-bein would directly give an ansatz for
  $F_{mnpq}$ without need to invert the metric. However, this leads to
  a more complicated expression involving $A_{mnp}$ and $A_{m_1 \dots
    m_6}$ contributions on the right-hand side.} Moreover, the
contributions from the generalised connection components do not
vanish. It is these terms that antisymmetrise the $\partial A$ terms
in $\partial \cV$ to give the field strength. Hence without these
terms the field strength components would not be fully
antisymmetric   --  a point that was noted in Ref.~\cite{NP}. We therefore
conclude that differentiating $A_{mnp}$ obtained from the non-linear
uplift flux ansatz is a simpler way of finding the internal components
of $F_{mnpq}$ than the ansatz derived from the internal GVP, see Eqn.~\eqref{eq:Fmnpq}.

While the generalised spin connection and non-metricity are
parametrised by $F_{mnpq}$ and $\fr$, the connections of the external
GVP \cite{Godazgar:2013dma} are given in terms of the $F_{\mu \nu \rho
  m}$ and $F_{\mu mnp}$ components of the 4-form field strength. In
E$_{7(7)}$ covariant form, the external GVP is \cite{Godazgar:2014sla}
\begin{equation} \label{exgvp}
  \partial_\mu \cV_{\cM \, AB} + 2 \hat{\cL}_{\cB_{\mu}} \cV_{\cM\,
    AB} + \cQ_{\mu}^{C}{}_{[A} \cV_{\cM\, B]C} = \cP_{\mu\, ABCD}
  \cV_{\cM}{}^{CD},
\end{equation}
where $\hat{\cL}$ is the E$_{7(7)}$ generalised Lie
derivative~\cite{Coimbra:2011ky, Berman:2012vc}~\footnote{The
  generalised Lie derivative encodes the diffeomorphisms and gauge
  transformations of the $D=11$ fields \cite{Coimbra:2011ky, Godazgar:2014sla}. In
  approaches where the base space is also enlarged, e.g.\
  Ref.~\cite{Berman:2012vc, Hohm:2013uia}, the partial derivatives
  also carry E$_{7(7)}$ indices.}
\begin{equation} \label{genLie} \hat{\cL}_{\Lambda} X_{\cM} \,=\,
  \Lambda^m \partial_m X_{\cM} \,+ \, 12 (t^{\bm{\alpha}})_{\cM}{}^{\cN}
  (t_{\bm{\alpha}})_{\cP}{}^{q8} \partial_{q} \Lambda^{\cP} X_{\cN}
\end{equation}
and the connection coefficients are of the form
\begin{align}
  \cQ_{\mu}^{A}{}_{B} &= - \textstyle{\frac{1}{2}} \Big[ {e^m}_a D_{m}
  B_{\mu}{}^{n} e_{n b} - ({e^p}_{a} \cD_{\mu} e_{p\, b}) \Big]
  \Gamma^{ab}_{AB} - \textstyle{\frac{\sqrt{2}}{12}}
  {e_{\mu}}{}^{\alpha} \left( F_{\alpha abc} \Gamma^{abc}_{AB} -
    \eta_{\alpha \beta \gamma
      \delta} F^{\beta \gamma \delta a} \Gamma_{a AB} \right), \\[3mm]
  \cP_{\mu ABCD}& = \textstyle{\frac{3}{4}} \Big[ {e^m}_a D_{m}
  B_{\mu}{}^{n} e_{n b} - ({e^p}_{a} \cD_{\mu} e_{p\, b}) \Big]
  \Gamma^{a}_{[AB} \Gamma^{b}_{CD]} - \textstyle{\frac{\sqrt{2}}{8}}
  {e_{\mu}}{}^{\alpha} F_{abc \alpha}
  \Gamma^{a}_{[AB} \Gamma^{bc}_{CD]} \notag \\[2mm]
  & \hspace{75mm} - \textstyle{\frac{\sqrt{2}}{48}} e_{\mu \, \alpha}
  \eta^{\alpha \beta \gamma \delta} F_{a \beta \gamma
    \delta}{\Gamma_{b}}_{[AB} \Gamma^{ab}_{CD]},
\end{align}
where
\begin{equation}
  \cD_\mu \equiv \partial_\mu - B_\mu{}^m D_m.
\end{equation}
We recall that $D_{m}$ is the covariant derivative with respective to
the connection $\Gamma^{p}_{mn}$ and $ e_{\mu}{}^{\alpha}$ is the vierbein.

Given a particular reduction ansatz, the external GVP \eqref{exgvp}
reduces to the Cartan equation of the scalars of the four-dimensional
maximal gauge theory~\cite{deWit:2007mt}:
\begin{equation} \label{4gvp}
  \partial_\mu \hat{\cV}_{\cM \, ij} - g A_{\mu}{}^{\cP} X_{\cP
    \cM}{}^{\cN} \hat{\cV}_{\cN\, ij}+ \cQ_{\mu}^{k}{}_{[i}
  \hat{\cV}_{\cM\, j]k}    = \cP_{\mu\, ijkl} \hat{\cV}_{\cM}{}^{kl},
\end{equation}
where $\hat{\cV}$ is given in Eqn.~\eqref{eq:cVhat} and $X_{\cM}$ are
generators of the gauge algebra and are related to the embedding
tensor $\Theta_{\cM}{}^{{\bm{\alpha}}}$ as follows
\begin{equation}
  X_{\cM} =\Theta_{\cM}{}^{\mathbf{{\bm{\alpha}}}} t_{\mathbf{{\bm{\alpha}}}}.
\end{equation}
The embedding tensor projects out at most 28 of the 56 vectors
$A_{\mu}{}^{\cP}$ \cite{deWit:2007mt}. The $\cQ_{\mu}^{i}{}_{j}$ are
related to $ \cQ_{\mu}^{A}{}_{B}$ by an inhomogeneous relation, while
$\cP_{\mu ijkl}$ are covariantly related to $\cP_{\mu ABCD} $ via the
eight Killing spinors of the vacuum solution of the maximal gauged
supergravity.

Let us consider the term proportional to $F_{\mu \nu \rho m}$ in
$\cP_{\mu \, ABCD}$. This term can be projected out as follows:
\begin{equation}
  \cP_{\mu \, ABCD} \cV^{n8 \; AB} \cV_{mn}{}^{CD} =
  \frac{\sqrt{2}}{8} e_{\mu \, \delta} \eta^{\alpha \beta \gamma
    \delta} F_{m \alpha \beta \gamma}.
\end{equation}
Thus, we obtain the uplift ansatz
\begin{equation} \label{eq:F31ansatz} F_{\mu \nu \rho m} = \frac{2
    \sqrt{2}}{3} i\, \eta_{\mu \nu \rho}{}^{\sigma} \left(
    \hat{\cV}^{\cM}{}_{ij} \partial_\sigma \hat{\cV}_{\cM \, kl} - g
    A_{\sigma}{}^{\cP} X_{\cP \cM}{}^{\cN} \hat{\cV}^{\cM}{}_{ij}
    \hat{\cV}_{\cN\, kl}\right) \cV^{n8 \; ij} \cV_{mn}{}^{kl}.
\end{equation}
This provides a non-linear ansatz for $F_{\mu \nu \rho m}$ for any
truncation of $D=11$ supergravity to four dimensions. Note that the
ans\"atze for $\cV^{n8} $ and $\cV_{mn}$ will be linear and follow
directly from the linear ans\"atze for the vectors.

In the $S^7$ truncation, the connections in Eqn.~\eqref{exgvp} and
\eqref{4gvp} are related via the eight Killing spinors $\eta^{i}$ on
the $S^7$~\cite{deWit:1986iy}
\begin{align}
  & \cQ_{\mu}^{i}{}_{j} = \eta^{i}_{A} \, \eta^{B}_{j} \left(
    \cQ_{\mu}^{A}{}_{B} - \frac{\sqrt{2} i}{4} \, m_7 \,
    A_{\mu}{}^{KL} K^{n \, KL} \eo_{n}{}^{a} \Gamma_{a}{}^{A}{}_{B}
  \right), \\[2mm]
  & \cP_{\mu \,ijkl} = \eta^{A}_{i} \, \eta^{B}_{j} \, \eta^{C}_{k} \,
  \eta^{D}_{l} \, \cP_{\mu \, ABCD}, \label{4dPmu}
\end{align}
where $A_{\mu}{}^{KL}$ are the 28 vectors of the $d=4$ theory that are
gauged. The generators of the gauge algebra are given
by~\cite{deWit:1982ig}
\begin{equation}
  X_{\cM \,  \cN}{}^{\cP} = 
  \begin{cases}
    X_{IJ \, KL }{}^{MN} = X_{IJ}{}^{KL}{}_{MN} = 2 \delta^{R[K}_{IJ}
    \delta^{L]R}_{MN} \\
    0 \qquad \textrm{otherwise}
  \end{cases}
\end{equation}
and the reduction ansatz for the relevant components of the 56-bein
are given in Eqns.~\eqref{eq:4d-vielbeinB} and
\eqref{eq:4d-vielbeinC}. With these substitutions,
Eqn.~\eqref{eq:F31ansatz} reduces to
\begin{align} \label{eq:F31ansatzs7}
  F_{\mu \nu \rho m} &= - \frac{\sqrt{2}}{48} \eta_{\mu \nu
    \rho}{}^{\sigma} K^{n \; IJ} K_{mn}{}^{KL} \left(u^{ij}{}_{IJ} +
    v^{ij IJ} \right) \left( u^{kl}{}_{KL}- v^{kl KL} \right)  \notag
  \\[3mm]
  & \hspace{10mm}\times\left( \hat{\cV}^{\cM}{}_{ij} \partial_\sigma
    \hat{\cV}_{\cM \, kl} - 2 \sqrt{2} m_7 A_{\sigma}{}^{MN}
    \hat{\cV}^{MP}{}_{ij} \hat{\cV}_{NP\, kl} - 2 \sqrt{2} m_7
    A_{\sigma}{}^{MN} \hat{\cV}^{MP}{}_{kl} \hat{\cV}_{NP\,
      ij}\right).
\end{align}
This is the non-linear uplift ansatz for $F_{\mu \nu \rho m}$ for the
$S^7$ reduction of $D=11$ supergravity. We note, as a check, that in
the SO$(7)^+$ sector the above expression reproduces the correct
result, \emph{viz.}
\begin{equation}
  F_{\mu \nu \rho m}= \frac{\sqrt{2}}{6} i\, \eta_{\mu \nu
    \rho}{}^{\sigma} \partial_{\sigma} \lambda \, \partial_m \xi. 
\end{equation}

The above ansatz for $F_{\mu \nu \rho m}$, \eqref{eq:F31ansatz},
provides a considerable simplification over computing the Hodge dual
of $F_{\mu m_1 \dots m_6}$ calculated using the ansatz for the metric,
3-form and 6-form. This is clear even in the relatively simple case of
the SO$(7)^+$ sector. The advantage of the ans\"atze
\eqref{eq:F31ansatz} for $F_{\mu\nu\rho m}$ (and its specialisation to
the $S^7$ reduction \eqref{eq:F31ansatzs7}) and \eqref{eq:f-explicit}
for the Freund-Rubin term is that they do not require differentiation
or the metric to be inverted.

The connection $\cP_{\mu \, ABCD}$ also depends on the $F_{\mu mnp}$
components of the field strength. However, as is the case with the
ansatz for $F_{mnpq}$, \eqref{eq:Fmnpq}, we do not obtain a direct
ansatz. Therefore, for the $F_{\mu mnp}$ and $F_{mnpq}$ components the
GVPs do not provide more efficient ans\"atze. However, these
components are easily calculated using the 3-form ansatz \cite{dWN13}.
We are fortunate that the GVPs give direct ans\"atze for the
components of the field strength that are otherwise difficult to
calculate.

The only remaining component of the field strength that we have not
thus far discussed is the $F_{\mu \nu mn}$ components, which does not
feature in the GVPs. However, this component does enter the fermion
supersymmetry transformations via
\begin{gather}
  \cG_{\alpha \beta AB} \equiv - \frac{1}{8} i\, \Delta^{-1/2}
  e_{[\alpha}{}^{\mu} e_{\beta]}{}^{\nu} \cD_{\mu} B_{\nu}{}^{n}
  \Gamma_{n AB} + \frac{\sqrt{2}}{32} i \Delta^{-1/2} F_{\alpha \beta
    mn} \Gamma^{mn}_{AB}. \label{cG}
\end{gather}
Comparing the fermion supersymmetry transformations in four
\cite{deWit:1982ig, deWit:2007mt} and eleven dimensions
\cite{deWit:1986mz}, we make the following identification
\begin{equation} \label{cH}
 \cH_{\alpha \beta\, ij} = 4 \sqrt{2}\, \eta^A_i \eta^B_j\,
 \cG_{\alpha \beta AB}, 
\end{equation}
where $\cH_{\alpha \beta\, ij}$ is related to the covariantised field
strength $\cG_{\alpha \beta}{}^{\cM}$ \cite{deWit:2007mt}
\begin{equation} \label{cHcG} \cH_{\alpha \beta\, ij} = \hat{\cV}_{\cM
    \, ij}\, \cG_{\alpha \beta}{}^{\cM}.
\end{equation}
Contracting Eqn.~\eqref{cG} with $\cV_{mn}{}^{AB}$ gives an imaginary
expression
\begin{equation}
  \cV_{mn}{}^{AB} \cG_{\alpha \beta AB} = -\frac{1}{8} i \Delta^{-1}
  F_{\alpha \beta mn} +\frac{3}{2} i \Delta^{-1} A_{mnp} \left(
    e^{\mu}{}_{[\alpha} e^{\nu}{}_{\beta]} \cD_{\mu} B_{\nu}{}^{p} -
    g^{pq} e^{\mu}{}_{[\alpha} \partial_{q} e_{\mu \, \beta]} \right). 
\end{equation}
Using Eqns.~\eqref{cH}, \eqref{cHcG} and the above equation, we obtain
the non-linear uplift ansatz for $F_{\mu \nu mn}$ for any reduction
\begin{equation} \label{eq:F22} F_{\mu \nu mn} = \sqrt{2} i
  \cV_{mn}{}^{ij} \hat{\cV}_{\cM \, ij} \cG_{\mu \nu}{}^{\cM} +
  \frac{3}{2} A_{mnp} \left( \cD_{[\mu} B_{\nu]}{}^{p} + g^{pq}
    e_{[\mu}{}^{\alpha} \partial_{q} e_{\nu] \, \alpha} \right).
\end{equation}
Specialising to the $S^7$ reduction gives
\begin{equation} \label{eq:F22s7} F_{\mu \nu mn} = \frac{\sqrt{2}}{8}
  \left(K_{mn}{}^{IJ} \cG_{\mu \nu\, IJ} - 12 \Delta^{-1} A_{mnp}
    K^{p}{}_{IJ} \cH_{\mu \nu}{}^{IJ} \right),
\end{equation}
where \cite{deWit:2007mt}
\begin{equation}
  \cG_{\mu \nu}{}^{\cM} = 
  \begin{pmatrix}
    \cH_{\mu \nu}{}^{IJ} \\[2mm]  \cG_{\mu \nu\, IJ}
  \end{pmatrix}.
\end{equation}
Hence, $F_{\mu \nu mn}$ is only non-trivial for four-dimensional
solutions with non-zero vector expectation values.

\section{General form of the Freund-Rubin term}
\label{sec:shell-conj-freund}

\subsection{The conjecture}
\label{sec:state-conjecture}

We observed in section \ref{sec:testing-non-linear} and appendices \ref{app:freund-rubin-so3t} and \ref{app:freund-rubin-su4} that for various
examples the Freund-Rubin term is proportional to the potential, with the
constant of proportionality given by $- m_7/(\sqrt{2}g^2)$
\cite{NP}, and a $y$-dependent part that contains variations of the
potential. Furthermore, the $y$-dependence only enters linearly via
the invariant scalars ($\xi$ in G$_2$ and $(\xi,\zeta)$ in
SO(3)$\times$SO(3), see appendix B). In particular, if the sector under consideration does not contain an
invariant scalar (such as SO(7)$^-$ or SU(4)$^-$), then $\fr$ is
$y$-independent and proportional to the potential. In the following,
we will state a general conjecture for the Freund-Rubin term that
respects all these observations.

First, we state the general expressions for the potential $V$ and its
variation $\delta V$ in terms of the tensors $u_{ij}{}^{IJ}$ and
$v_{ij\;IJ}$. We define the $T$-tensor \cite{deWit:1982ig}
\begin{equation}
  \label{eq:T}
  T_i{}^{jkl} =
  \left(
    u^{kl}{}_{IJ} + v^{kl\;IJ}
  \right)
  \left(
    u_{im}{}^{JK} u^{jm}{}_{KI} - v_{im\;JK} v^{jm\;KI}
  \right)
\end{equation}
and its components
\begin{equation}
  \label{eq:A1-A2}
  A_1^{ij} = \frac{4}{21} T_k{}^{ikj},\qquad A_{2\;i}{}^{jkl} =
  -\frac{4}{3} T_i{}^{[jkl]}.
\end{equation}
In terms of the above tensors the potential is given by
\cite{deWit:1982ig}
\begin{align} \label{eq:potential}
  V \,=\, & \frac{1}{24}g^2  A_{2\;i}{}^{jkl} A_2{}^i{}_{jkl}  -
  \frac{3}{4}g^2 A_1^{ij} A_{1\, ij} \, .
\end{align}

In order to determine the variation of the potential, we consider an
infinitesimal E$_{7(7)}$ variation of the 56-bein of the form
\cite{deWit:1983gs}
\begin{equation}
  \delta \cV = - \frac{\sqrt{2}}{4} 
  \begin{pmatrix}
    0 & \Sigma^{ijkl} \\
    \Sigma_{ijkl} & 0 
  \end{pmatrix} \cV,
\end{equation}
where $\Sigma$ is complex selfdual. Given the variation of the 56-bein
given above, to first order, the potential varies
as~\cite{deWit:1983gs}
\begin{align} \label{eq:varpot} \delta V =& \frac{\sqrt{2}}{24}g^2
  Q^{ijkl} \Sigma_{ijkl} + \mathrm{h.c.},
\end{align}
where the $Q$-tensor is
\begin{equation}
  \label{eq:Qijkl}
  Q^{ijkl} = \frac{3}{4} A_{2\;m}{}^{n[ij} A_{2\;n}{}^{kl]m} -
  A_1{}^{m[i} A_{2\;m}{}^{jkl]}.
\end{equation}
Since, the expression on the right-hand side of Eqn.~\eqref{eq:varpot}
gives the variation of the potential to first order, it must vanish at
the stationary points. In particular, since $\Sigma_{ijkl}$ is an
arbitrary complex selfdual tensor, $Q^{ijkl}$ is complex anti-selfdual
at stationary points.

We define a complex selfdual combination of $u$ and $v$ tensors
\begin{equation}\label{Sigmahat}
  \hat \Sigma_{ijkl} (x,y) \,\equiv \, 
  \left(
    u_{ij}{}^{IJ}(x) u_{kl}{}^{KL}(x) - v_{ij\;IJ}(x) v_{kl\;KL}(x)
  \right) K^{IJKL}(y),
\end{equation}
where we have written out the coordinate dependence explicitly so as
to make the dependence of $\hat\Sigma$ on all eleven coordinates
clear. Making use of the $Q$-tensor, we are now able to formulate a
conjecture for the Freund-Rubin term:
\begin{equation}
  \label{eq:conjecture}
  \fr = -\frac{m_7}{\sqrt{2}g^2}
  \left(
    V - \frac{g^2}{24}
    \left(
      Q^{ijkl} \hat \Sigma_{ijkl} + \mathrm{h.c.}
    \right)
  \right).
\end{equation}
The second term on the right-hand side is inevitably $y$-dependent,
and it vanishes when $Q^{ijkl}$ is complex {\em anti}-selfdual, which
is precisely the minimisation condition for the potential.

To prove this formula, one has to manipulate
Eqn.~(\ref{eq:f-explicit}) using E$_{7(7)}$ identities for the $u$ and
$v$ tensors \cite{deWit:1982ig, deWit:1986iy}. However, the proof will
also probably require identities derived from the quartic invariant
(see, e.g.\ Ref.~\cite{so(8)}). We leave this proof (which is probably
even more complicated than the one given in Ref.~\cite{deWit:1986iy}
for the $y$-independence of the $A_1$ and $A_2$ tensors coming from
the $S^7$ truncation) for future work. In the remainder of this
section, we will prove the conjecture up to quadratic order and
verify it for the G$_2$ invariant sector.

\subsection{Proof of the conjecture up to quadratic order}
\label{sec:test-conj-up}

In this section, we prove the equality of Eqns.~\eqref{eq:f-explicit}
and \eqref{eq:conjecture} for a perturbative expansion of the $u$ and
$v$ tensors. As in Eqn.~\eqref{G2vielbein} we use the unitary gauge,
\begin{equation}
  \label{eq:3}
  \mathcal{V} =
  \exp  \begin{pmatrix}
    0 & \phi_{IJKL} \\[1mm]  \phi^{IJKL} & 0
  \end{pmatrix},
\end{equation}
where we do not need to distinguish between SU(8) and SO(8) indices.
Thus, 
\begin{equation}
  u_{IJ}{}^{KL} = (\cosh \phi)_{IJ}{}^{KL}, \qquad v_{IJKL}
  = (\sinh \phi)_{IJKL}.
\end{equation}
Here, we denote
\begin{gather}
  (\phi^0)_{IJ}{}^{KL} = \delta_{IJ}{}^{KL}, \qquad
  (\phi^2)_{IJ}{}^{KL} = \phi_{IJMN}\phi^{MNKL}.
\end{gather}
Complex conjugation is realised by raising and lowering indices.
Furthermore, the potential is complex selfdual,
\begin{equation}
  \phi_{IJKL}^* = \phi^{IJKL} = \frac{1}{24} \epsilon^{IJKLMNPQ}
  \phi_{MNPQ}.
\end{equation}
Up to quadratic order, we obtain
\begin{equation}
  u_{IJ}{}^{KL} \,=\, \delta^{KL}_{IJ} \,+\,  \frac{1}{2} \phi_{IJMN}
  \phi^{MNKL} \,+\, \cO(\phi^4)\; , 
  \qquad v_{IJKL} \,=\,  \phi_{IJKL} \, + \, \cO(\phi^3)\,  .
\end{equation}

Substituting the expansions for the $u$ and $v$ tensors in the
expressions for $\rX_{rs}{}^{ijkl}$ and $\rY_{rs}{}^{ijkl}$,
\eqref{eq:X} and \eqref{eq:Y}, we find up to terms  $\cO(\phi^2)$,
\begin{align}
  \rX_{rs}{}^{ijkl} \rX^{rs}{}_{ijkl} &= 168 + 19 \, \phi^{IJKL}
  \phi_{IJKL} - \phi_{IJKL} \phi_{IJKL} \notag \\ 
  & \hspace{20mm}+ 3 \, K^{IJKL} \left(2 \, \phi_{IJKL} + 3 \,
    \phi^{IJMN} \phi_{MNKL} \right) + \frac{1}{24} \left( K^{IJKL}
    \phi_{IJKL} \right)^2, \\[2mm]
  \rX_{rs}{}^{ijkl} \rY^{rs}{}_{ijkl} &= - 6 \, \phi^{IJKL} \phi_{IJKL}
  - 6\, \phi_{IJKL} \phi_{IJKL} \notag \\ 
  & \hspace{20mm} + 2 \, K^{IJKL} \left(4 \, \phi_{IJKL} + 3 \,
    \phi^{IJMN} \phi_{MNKL} \right) + \frac{1}{4} \left( K^{IJKL}
    \phi_{IJKL} \right)^2, \\[3mm] 
  \rY_{rs}{}^{ijkl} \rY^{rs}{}_{ijkl} &= 32 \, \phi^{IJKL} \phi_{IJKL}
  + 24 \, K^{IJKL} \phi^{IJMN} \phi_{MNKL},
\end{align}
where now all the $y$-dependence is contained in $K^{IJKL}(y)$.
In deriving the above expressions, we make use of the following
identities
\begin{align}
  \label{eq:K-Identitites}
  K^{IJKP} K_{LMNP} &= 6 \delta^{IJK}_{LMN} + 9 \delta^{[I}_{[L}
  K^{JK]}{}_{MN]},\\
  K^{[IJKL} K^{M]NPQ} &= \frac{1}{5} \epsilon^{IJKLMNPQ} + 12
  K^{[IJK}{}_{[N} \delta^{L}{}_{P} \delta^{M]}{}_{Q]},\\
  K^{m\,IJ} K^{n\,KL} K_{mn}{}^{MN} &= \, 8 \delta^{[I}{}_{[K}
  \delta^{J][M} \delta^{N]}{}_{L]} + 4 \delta^{[M}{}_{[I}
  K^{N]}{}_{J]KL} + 4\delta^{[K}{}_{[M} K^{L]}{}_{N]IJ} -
  4\delta^{[I}{}_{[K} K^{J]}{}_{L]MN}.
\end{align}
It is now straightforward to show that, up to quadratic order, the
Freund-Rubin term, \eqref{eq:f-explicit}, is
\begin{equation}
  \fr = \sqrt{2} m_7 \left( 3 + \frac{1}{12}
    \phi_{IJKL} K^{IJKL} + \frac{1}{6} \phi_{IJKL} \phi^{IJKL}
  \right) + \mathcal{O}(\phi^3).
\end{equation}

We also find that
\begin{gather}
  V/g^2 = - 6 - \frac{1}{3} \phi_{IJKL} \phi^{IJKL} +
  \mathcal{O}(\phi^3), \qquad Q^{ijkl} \hat \Sigma_{ijkl} = 2
  \phi_{IJKL} K^{IJKL} + \mathcal{O}(\phi^3).
\end{gather}
Thus it is easy to verify that the conjectured expression,
\eqref{eq:conjecture}, reproduces the expression for the Freund-Rubin
term up to quadratic order in the scalar expectation values.

\subsection{Testing the conjecture in the G$_2$ invariant sector}
\label{sec:test-conj-g_2}

At the stationary points, it has already been established that the
conjecture (\ref{eq:conjecture}) holds for the G$_2$ invariant sector
\cite{NP}, see Eqns.~\eqref{eq:f-G2} and \eqref{fr:so7}. Therefore, it just remains to
prove that the $y$-dependent parts of Eqn.~(\ref{eq:f-G2-explicit})
and Eqn.~(\ref{eq:conjecture}) coincide, \emph{viz.}
\begin{equation}
  \label{eq:conjecture-test}
  \left(
    Q^{ijkl} \hat \Sigma_{ijkl} + h.c.
  \right)
  = 16 \xi (c + vs)^2 c v s
  \left(
    3 v s + 2 v^3 s^3 - 3 c - c v^2 s^2 - c^2 v s
    + 2 c^3
  \right),
\end{equation}
where again all the $y$-dependence is contained in the factor
$\xi(y)$.

Equation~\eqref{eq:Qijkl} provides an expression for the $Q$-tensor in
terms of the $u$ and $v$ tensors with four free SU(8) indices. Thus, we can use
Eqns.~(\ref{eq:u-G2}) and (\ref{eq:v-G2}) to write the $Q$-tensor in
terms of contracted G$_2$ invariant tensors, \eqref{eq:G2-basis}, with
four free SO(8) indices
\begin{equation}
  Q^{ijkl} \to Q^{IJKL}.  
\end{equation}
In this case, unlike in section \ref{sec:freund-rubin-factor}, the
$u,v$ tensors are not necessarily contracted over index pairs.
However, the resulting expression for $Q^{IJKL}$ must be G$_2$
invariant. Hence, we should be able to write it in the basis given in
Eqn.~(\ref{eq:G2-basis}). In particular, it is totally antisymmetric,
so we must find
\begin{equation}
  \label{eq:pre-Q}
  Q^{IJKL} = c_+(\lambda,\alpha) C_+^{IJKL} + c_-(\lambda,\alpha)
  C_-^{IJKL}
\end{equation}
for some functions $c_\pm$.

An efficient way to work out the contractions of SO(8) indices in
$Q^{IJKL}$ is to use the SO(7) decomposition of the G$_2$ invariants
(\ref{eq:G2-basis}). An SO(8) index decomposes as $I = (i,8)$, where
$i$ is an SO(7) index that runs from 1 to 7. The decomposition of
$C_\pm^{IJKL}$ is \cite{deWit:1983gs}
\begin{equation}
  C_\pm^{ijk8} = C^{ijk}, \qquad C_\pm^{ijkl} = \mp \frac{1}{6} \eta'
  \epsilon^{ijklmnp} C_{mnp},
\end{equation}
with an arbitrary phase $\eta'$. This phase will drop out in our
calculations. The SO(7) tensor $C^{mnp}$ satisfies \cite{deWit:1983gs}
\begin{equation}
  \label{eq:SO(7)-Ids}
  C^{[mnp} C^{q]rs} = -\frac{1}{4} \eta' \epsilon^{mnpq[r}{}_{tu}
  C^{s]tu}, \qquad C^{mnr}C_{pqr} = 2\delta^{mn}_{pq} - \frac{1}{6}
  \eta' \epsilon^{mn}{}_{pqrst} C^{rst}.
\end{equation}
Moreover, the $D_-$-tensor decomposes as follows:
\begin{equation}
  D_-^{ijkl} = D_-^{i8k8} = 0, \qquad D_-^{ijk8} = - D_-^{k8ij} = 4
  C^{ijk} \qquad \Rightarrow \quad D_-^{[IJ\,KL]} = 0.
\end{equation}
For $D_+^{IJKL}$, we find the convenient SO(8) property
\begin{equation}
  D_+^{IJ}{}_{KL} = \frac{2}{3} D_+^{M[I}{}_{M[K} \delta^{J]}{}_{L]},
  \qquad \Rightarrow \quad D_+^{[IJ\,KL]} = 0
\end{equation}
so we only need
\begin{equation}
  D_+^{Mi}{}_{Mj} = -6 \, \delta^i_j, \qquad D_+^{M8}{}_{M8} = 42.
\end{equation}

Using all these SO(7) decompositions together with the identities for
the $C$-tensor in Eqn.~(\ref{eq:SO(7)-Ids}), we find exactly the
anticipated form, Eqn.~(\ref{eq:pre-Q}) with
\begin{align}
  \label{eq:c+}
  c_+(\lambda,\alpha) =& \frac{1}{2} c^3vs \left( 5c - 4 c^3 + 2 s^2
    \sin^2 \alpha \right) - \frac{3}{2} c^2 v^2 s^4 \sin^2 \alpha
  - \frac{1}{2} c^2v^3s^3 (1-c)(3+7c) \sin^2 \alpha \nonumber\\[2mm]
  &+ \frac{1}{2} v^4 s^4 \left(s^2 (5 - 3c) \sin^2 \alpha - c - 4
    c^3\right) + 2 v^6s^6 (c - 1) \sin^2 \alpha + \frac{3}{2} i v s^2
  \left( c^3 + v s^3 - c^2\right) \sin^3 \alpha
  \\[2mm]
  \label{eq:c-}
  c_-(\lambda,\alpha) =& - \frac{3}{2} s^2 \left(v^2c^3 + v^3s^3 + c^2
    \sin^2 \alpha \right) \sin^2 \alpha + \frac{1}{2} i c^4 s
  \sin\alpha \left(4 s^2 - 1 \right) \nonumber\\[2mm]
  &+ \frac{1}{2} i c^2 v^2 s^3 (3 +2c - 7 c^2) \sin^3 \alpha +
  \frac{1}{2} i c v^3 s^4 \left(5 + 7 s^2 \sin^2 \alpha - 3c
  \right) \sin \alpha \nonumber\\[2mm]
  & - i c^3 v^4 s^3 \sin \alpha + \frac{5}{2} i v^5 s^6 \sin \alpha +
  2 i v^7 s^6 (c-1) \sin \alpha.
\end{align}
At the G$_2$ invariant stationary point,
\begin{equation}
  c^2 = \frac{2\sqrt{3} + 3}{5}, \qquad s^2 = \frac{
    2\sqrt{3}-2}{5}, \qquad v^2 = \frac{3 - \sqrt{3}}{4},
\end{equation}
the $Q$-tensor is indeed, complex anti-selfdual because $c_+$ becomes
purely imaginary and $c_-$ purely real.

We compute $\hat \Sigma_{IJKL} C_\pm^{IJKL}$ using the above SO(7)
decomposition. Identifying,
\begin{equation}
  C_+^{IJKL} K^{IJKL} = -16\xi, \qquad C_-^{IJKL} K^{IJKL} = 0,
\end{equation}
we find
\begin{equation}
  \label{eq:Selfdual-G2}
  \hat \Sigma_{IJKL} C_+^{IJKL} = - 16\xi (c \sin^2\alpha +
  \cos^2\alpha), \qquad \hat \Sigma_{IJKL} C_-^{IJKL} = 8 i \xi
  \sin2\alpha (c-1).
\end{equation}
As expected, these expressions are linear in the invariant scalar
$\xi$. Eqn.~(\ref{eq:conjecture-test}) then follows immediately from
Eqns.~(\ref{eq:pre-Q},\ref{eq:c+},\ref{eq:c-}) and
Eqn.~(\ref{eq:Selfdual-G2}).

\section{Outlook}
\label{sec:outlook}

In this paper, we derive an explicit formula for the Freund-Rubin
term, \eqref{eq:f11d}, for any consistent truncation of $D=11$
supergravity to four dimensions by means of the internal generalised
vielbein postulate \cite{Godazgar:2013dma}. In the case of the $S^7$
reduction this reduces to \eqref{eq:f-explicit}. Previously,
the Freund-Rubin term could be computed using the uplift ans\"atze for the
6-form and 3-form, which involves inverting the metric and
differentiating. The new formulae are much simpler. Moreover, for the
$S^7$ truncation, we conjecture that the Freund-Rubin term is given by
the potential for the scalars of the truncated $d=4$ supergravity and
a variation of the potential. While the corresponding on-shell
conjecture has already been in the literature \cite{NP}, we propose a
formula, \eqref{eq:conjecture}, that bears this conjecture out more
concretely (off-shell). A corollary of our conjecture is that for
sectors that are purely characterised by pseudoscalar expectation
values, the Freund-Rubin term is $y$-independent and is completely
given by the scalar potential. We prove the conjecture up to quadratic
order in the scalar expectation value and verify it for the G$_2$
invariant sector. In the future, we hope to provide a proof of this
conjecture.

The GVPs and fermion supersymmetry transformations provide a new vista
on the form of the $D=11$ field strength that arises from uplifting
$d=4$ solutions. Given the striking simplicity of the conjectured
Freund-Rubin term, a natural question that we can now investigate,
arises: do the other components of the field strength take a similarly
simple form that depend on very general data of the reduced theory,
such as the scalar potential or its derivatives. Another aspect that
we would like to investigate is whether the conjectured form of the
Freund-Rubin term holds in general for all truncations of any theory.
A setting in which the analogous question can be addressed using
similar methods (analysis of GVPs and fermion supersymmetry
variations) is the reduction of type IIB supergravity to five
dimensions, where the necessary framework exists
\cite{Ciceri:2014wya}---nonlinear ans\"atze, which arise from an
analysis of the supersymmetry transformations of the vectors
\cite{Ciceri:2014wya}, have been proposed \cite{Lee:2014mla} and
presented explicitly \cite{Baguet:2015sma} in this case. Furthermore, in this case, the analysis of the supersymmetry transformations of the vectors has already been used by Pilch and Warner (appendix A of Ref.~\cite{Pilch:2000ue}) to derive uplift formulae for the metric and the dilaton.

Our study of reductions of $D=11$ supergravity to four dimensions
shows that consistent truncations seem to have simple, generic
features that are obscured by the complexity of particular examples.
With duality symmetry as a guide \cite{deWit:1986mz,
  Godazgar:2013dma}, we are able to tease out these features and it is
hoped that in the future we will learn something very general and
conceptually deep about \emph{all} reductions.

\section*{Acknowledgments}
We would like to thank Gary Gibbons, Krzysztof Pilch and Chris Pope
for useful discussions. H.G.\ and M.G.\ would like to thank the AEI,
in particular H.N., as well as the Mitchell Institute, TAMU, in
particular Chris Pope, for their generous hospitality. H.G.\ and M.G.\
are supported by King's College, Cambridge. H.G.\ acknowledges funding
from the European Research Council under the European Community's
Seventh Framework Programme (FP7/2007-2013) / ERC grant agreement no.
[247252].

\appendix

\section{Contractions of G$_2$ invariants with Killing forms}
\label{sec:contr-g_2-invar}

The G$_2$ invariant tensors can be used to define the following
tensors on the round $S^7$ \cite{deWit:1984nz}
\begin{gather}
  \xi_m =\; \frac{1}{16} C_+^{IJKL} K_{mn}^{IJ} K^{n \, KL}, \qquad
  \xi_{mn} =\; -\frac{1}{16} C_+^{IJKL} K_{m}^{IJ} K_n^{KL}, \qquad
  \xi =\; \mathring g^{mn} \xi_{mn}, \notag \\[2mm]
  \mathring S_{mnp} =\; \frac{1}{16} C_-^{IJKL}
  K_{[mn}^{IJ}K_{p]}^{KL}. \label{eq:XiS}
\end{gather}

We write terms like e.g. $D_-^{IJKL} K_m^{IJ} K_{np}^{KL}$ in terms of
the $S^7$ tensors in \eqref{eq:XiS}. These fulfill the identities
\begin{gather}
  \label{eq:XiSIdentities}
  \xi_m \xi_n = (9 - \xi^2) \mathring g_{mn} - 6(3 - \xi) \xi_{mn},
  \qquad \xi_m \xi^m = (21 + \xi)(3 - \xi) ,\\[2mm]
  S^{mnr} S_{pqr} = 2 \delta^{mn}_{pq} + \frac{1}{6} \mathring
  \eta^{mn}{}_{pqrst} S^{rst}, \ S^{[mnp} S^{q]rs} = \frac{1}{4}
  \mathring \eta^{mnpq [r}{}_{tu} S^{s]tu}, \ S^{m[np} S^{qr]s} =
  \frac{1}{6} \mathring \eta^{npqr(m}{}_{tu} S^{s)tu}.
\end{gather}

Together with the inverse relations of Eqn.~(\ref{eq:XiS}),
\begin{equation}
  C_+^{IJKL} = \frac{1}{6}\xi K_m^{[IJ} K^{m \, KL]} -
  \frac{3}{2} \xi^{mn} K_m^{[IJ} K_n^{KL]} + \frac{1}{12} \xi^m
  K_{mn}^{[IJ} K^{n \, KL]}, \quad
  S_{mnp} = \frac{1}{16} C_-^{IJKL} K_{[mn}^{[IJ} K_{p]}^{KL]},
\end{equation}
we obtain 
\begin{gather}
  \delta^{IJ}_{KL} K^{m\;IJ}K_n^{KL} = 8 \delta^m_n , \qquad
  \delta^{IJ}_{KL} K_{mn}^{IJ}K^{p\;KL} = 0, \qquad \delta^{IJ}_{KL}
  K^{mn\;IJ}K_{pq}^{KL} = 16 \delta^{mn}_{pq},
  \nonumber\\[5pt]
  C_+^{IJKL} K_m^{IJ}K_n^{KL} = -16\xi_{mn}, \qquad C_+^{IJKL}
  K_{mn}^{IJ}K^{p\;KL} = \frac{16}{3} \xi_{[m} \delta_{n]}{}^p, \nonumber\\[5pt]
  C_+^{IJKL} K^{mn\;IJ}K_{pq}^{KL} = \frac{32}{3} \xi \delta^{mn}_{pq}
  - 64 \xi^{[m}{}_{[p} \delta^{n]}{}_{q]}, \nonumber\\[5pt]
  C_-^{IJKL} K_m^{IJ}K_n^{KL} = 0, \qquad C_-^{IJKL}
  K_{mn}^{IJ}K_p^{KL} = 16 S_{mnp}, \qquad C_-^{IJKL}
  K_{mn}^{IJ}K_{pq}^{KL} = -\frac{8}{3} \mathring
  \eta_{mnpqrst} S^{rst}, \nonumber\\[5pt]
  D_+^{IJKL} K_m^{IJ}K_n^{KL} = 0, \qquad D_+^{IJKL}
  K_{mn}^{IJ}K_p^{KL} = - 48 \xi^q{}_{[m} S_{np]q} + \frac{16}{3} \xi
  S_{mnp} + \frac{4}{9} \mathring \eta_{mnpqrst}
  \xi^q S^{rst}, \nonumber\\[5pt]
  D_+^{IJKL} K_{mn}^{IJ}K_{pq}^{KL} = \frac{32}{3} \xi_{[m} S_{npq]}
  -8 \mathring \eta_{mnpqrst}\xi_{u}{}^r S^{stu} + \frac{8}{9} \xi
  \mathring \eta_{mnpqrst} S^{rst}, \nonumber\\[5pt]
  D_-^{IJKL} K_m^{IJ}K_n^{KL} = \frac{16}{3} S_{mnp} \xi^p, \nonumber\\[5pt]
  D_-^{IJKL} K_{mn}^{IJ}K_p^{KL} = - 32 \xi^q{}_{[m} S_{n]pq} + 16
  S_{mnq}\xi^q{}_p + \frac{16}{3} \xi S_{mnp} - \frac{4}{9}
  \mathring \eta_{mnpqrst} \xi^q S^{rst} , \nonumber\\[5pt]
  D_-^{IJKL} K_{mn}^{IJ}K_{pq}^{KL} = \frac{16}{3} \xi_{[m}S_{n]pq} +
  \frac{16}{3} S_{mn[p} \xi_{q]} - \frac{16}{3} \mathring
  \eta_{mn[p|rstu} \xi^r{}_{q]} S^{stu} + \frac{16}{3} \mathring
  \eta_{[m|pqrstu} \xi^r{}_{n]} S^{stu}.
\end{gather}

\section{Freund-Rubin term in the SO(3)$\times$SO(3) invariant sector}
\label{app:freund-rubin-so3t}

The SO(3)$\times$SO(3) invariant sector is given by 
\begin{equation}
  \phi_{IJKL} \,=\, \frac{\lambda}{2} \left[ \cos \alpha\, \left(
      Y_+^{IJKL}  \,+\, i \, Y_-^{IJKL} \right) - \sin\alpha \, \left(
      Z_+^{IJKL}  \,-\, i \, Z_-^{IJKL} \right) \right],
\end{equation}
where $Y_{\pm}$ and $Z_{\pm}$ are SO(3)$\times$SO(3) invariant
tensors.

The scalar potential reads
\begin{equation}
  V(\lambda) = \frac{g^2}{2}(\tilde{s}^4 - 8 \tilde{s}^2 - 12).
\end{equation}
Here, $\tilde{s} = \sinh \sqrt{2} \lambda$ and $\tilde{c} = \cosh \sqrt{2} \lambda$.
Note that $V$ does not depend on $\alpha$ \cite{Bobev:2011rv}.

In Ref.~\cite{Godazgar:2014eza}, the $u$ and $v$ tensors are given in
terms of SO(3)$\times$SO(3) invariants
\begin{gather}
  Y_+^{IJKL}, Z_+^{IJKL}: \mathrm{selfdual}, \qquad Y_-^{IJKL},
  Z_-^{IJKL}: \mathrm{anti-selfdual},\\ \Pi^{IJKL} = \frac{1}{8}
  \left(Y_+^{IJMN} + i Y_-^{IJMN}\right)\left(Y_+^{MNKL} - i
    Y_-^{MNKL}\right)
\end{gather}
from which we define the following $y$-dependent scalars
\begin{gather}
  \xi (y) =\; -\frac{1}{16} Y^+_{IJKL} K_{m}^{IJ} K^{ m \,KL}, \quad
  \zeta (y) =\; -\frac{1}{16} Z^+_{IJKL} K_{m}^{IJ} K^{m \,KL}.
\end{gather}

Using the results in Ref.~\cite{Godazgar:2014eza} for the $u$ and $v$
tensors and identities stated in that paper, we find
\begin{align}
  \fr =& \frac{m_7}{\sqrt{2}}\left(6 + 4 \tilde{s}^2 - \frac{\tilde{s}^4}{2}\right) +
  \frac{m_7}{6}(\zeta \sin \alpha - \xi \cos
  \alpha )( 4 \tilde{s} \tilde{c} - \tilde{s}^3 \tilde{c}) \notag \\[3mm]
  =& \frac{m_7}{\sqrt{2} g^2} \left( -V(\lambda) - \frac{1}{12} \left(
      \zeta \sin \alpha - \xi \cos \alpha \right)
    \frac{\mathrm{d}V}{\mathrm{d}\lambda} \right).
\end{align}
Again the $y$-dependence is contained in $\xi$ and $\zeta$.

\section{Freund-Rubin term in SU(4)$^-$ invariant sector}
\label{app:freund-rubin-su4}

The SU(4)$^-$ invariant sector is parametrised by a single
pseudoscalar expectation value,
\begin{equation}
  \phi_{IJKL} \,=\, \frac{1}{2} \, i \lambda Y_{-}^{IJKL}.
\end{equation}
In this case, we find that
\begin{equation}
  \fr = - \sqrt{2} m_7 c^2  \left( c^2 - 4 \right)
\end{equation}
for $c = \cosh 2\lambda$. We note that, since this sector only
contains a pseudoscalar, i.e.\ there are no selfdual tensors, the
Freund-Rubin term is indeed $y$-independent even away from the
stationary point.

\bibliographystyle{utphys}
\bibliography{literature}

\end{document}